\documentclass[a4,useAMS,usenatbib,usegraphicx]{mn2e}

\def\kms{\,km\,s$^{-1}$}

\def\apj{ApJ}
\def\apjl{ApJL}
\def\mnras{MNRAS}
\def\pasp{PASP}
\def\aj{AJ}

\def\aap{A\&A}
\def\apjs{ApJS}

\def\kms{\,km\,s$^{-1}$}

\def\apj{ApJ}
\def\apjl{ApJL}
\def\mnras{MNRAS}
\def\pasp{PASP}
\def\aj{AJ}

\def\aap{A\&A}
\def\apjs{ApJS}

\usepackage{lscape}
\usepackage{natbib}
   \title[AGES: II Abell 1367 and its outskirts]{The Arecibo Galaxy Environment Survey: II. A HI view of the Abell cluster 1367 and its outskirts.}

  \author[L. Cortese et al.]{L. Cortese$^{1}$, R. F. Minchin$^{2}$, R. R. Auld$^{1}$, J. I. Davies$^{1}$, B. Catinella$^{2,3}$, E. Momjian$^{2}$,  \newauthor  J. L. Rosenberg$^{4}$, R. Taylor$^{1}$, G. Gavazzi$^{5}$, K. O'Neil$^{6}$, M. Baes$^{7}$, A. Boselli$^{8}$, \newauthor G. Bothun$^{9}$, B. Koribalski$^{10}$, S. Schneider$^{11}$, W. van Driel$^{12}$\\
$^1$ School of Physics and Astronomy, Cardiff University, Cardiff CF24 3AA, UK\\
$^2$ NAIC-Arecibo Observatory, HC3 Box 53995, Arecibo, PR 00612, USA\\
$^3$ Max-Planck-Institut f\"{u}r Astrophysik, D-85748 Garching, Germany.\\
$^4$ Harvard-Smithsonian Centre for Astrophysics, 60 Garden Street, MS 65, Cambridge, MA 02138-1516, USA\\
$^5$ Universit\'{a} degli Studi di Milano-Bicocca, Piazza della Scienza 3, 20126 Milano, Italy \\
$^6$ National Radio Astronomy Observatory, Green Bank, WV 24944, USA\\
$^7$ Sterrenkundig Observatorium, Universiteit Gent, Krijgslaan 281 S9, B-900 Gent, Belgium\\
$^8$ Laboratoire d'Astrophysique de Marseille, BP8, Traverse du Siphon, F-13376 Marseille, France\\
$^9$ Physics Department, University of Oregon, 1371 East 13th Avenue, Eugene, OR 97403, USA\\
$^{10}$ Australia Telescope National Facility, CSIRO, Epping, NSW, Australia\\
$^{11}$ Department of Astronomy, University of Massachusetts, Amherst, MA 01003, USA\\
$^{12}$ Observatoire de Meudon, 5 Place Jules Janssen, 92195 Meudon, France\\
    }

\begin{document}
\date{Accepted 2007 October 28.  Received 2007 October 1; in original form 2007 August 10 }
\pagerange{\pageref{firstpage}--\pageref{lastpage}} \pubyear{2007}

\maketitle

\label{firstpage}

\begin{abstract}
We present 21 cm HI line observations of 5$\times$1 square degrees centered on the local Abell cluster 1367 obtained as 
part of the Arecibo Galaxy Environment Survey. 
One hundred sources are detected (79 new HI measurements and 50 new redshifts), more than half 
belonging to the cluster core and its infalling region.
Combining the HI data with SDSS optical imaging we show that our HI selected sample follows 
scaling relations similar to the ones usually observed in optically selected samples. 
Interestingly all galaxies in our sample appear to have nearly the same baryon fraction independently 
of their size, surface brightness and luminosity.
The most striking difference between HI and optically selected samples resides in their large scale distribution: 
whereas optical and X-ray observations trace the cluster core very well, in HI 
there is almost no evidence of the presence of the cluster. 
Some implications on the determination of the cluster luminosity function and HI distribution 
for samples selected at different wavelength are also discussed. 
\end{abstract}


\section{Introduction}
With the advent of wide field surveys our knowledge 
of galaxy properties has significantly improved.
The large amount of high quality data of nearby and high redshift objects obtained so far 
is allowing astronomers to shed light on how galaxies formed and 
evolved at different redshifts and in different environments (e.g. \citealp{cowie96,phenomen,kauff03}).
However our picture of the Universe still remains limited since it is mostly based on 
the study of rest-frame optical selected objects.  
First of all optically-selected galaxies already contain, by definition, a large amount of stars which may
not be the case in younger or less evolved galaxies. Secondly, surface brightness selection is
very dramatic and nearly all catalogued objects lie in an extremely narrow range of
surface brightnesses \citep{freeman70}.\\ 
Since each wavelength has its own selection effects and traces different baryonic components, the combination of blind surveys at various frequencies is of vital importance. Alternative 
methods of baryon selection will allow us to have a less 
biased view of our Universe and enable us to correctly reconstruct the evolutionary history of galaxies (e.g. \citealp{vero06}).
In particular, the most interesting results can come from the comparison of optical surveys with 
blind surveys tracing components not directly related to 
stellar emission like, for example, neutral and molecular hydrogen.
For molecular hydrogen we will have to wait a couple of years for 
the first results from ALMA, but
with the advent of multi-beam instruments on large single dish radio telescopes it has 
become possible during the last decade to carry out fully sampled 
blind HI surveys of the sky (e.g. \citealp{dualbeam,henning00,barnes01,lang03,minchin03,davies04}). 
Neutral hydrogen is extremely important because 
it represents the fuel for the future star formation activity of a galaxy. It is also one of the galactic 
components that can most easily be affected by environmental mechanism (e.g. ram pressure, tidal interaction) influencing 
its star formation history.
Therefore HI line observations of galaxies can provide us with some of
the most powerful diagnostics on the role of the environment
in regulating the evolution of galaxies \citep{haynes,giova85}. What we lack is 
deep and high spatial resolution HI blind surveys of cluster of galaxies 
and their outskirts.
The HI Parkes all Sky Survey (HIPASS) has represented a real breakthrough for HI blind surveys, but
its low spatial resolution ($\sim$15 arcmin) has made it very difficult to carry out detailed studies 
of high density regions.
Fortunately, with the refurbishing of the 305m Arecibo radio telescope, blind HI surveys 
with much improved sensitivity, spatial  (3.5 arcmin $\rm beam^{-1}$) and  velocity (5 $\rm km~s^{-1}$) resolution have become possible 
thanks to the installation of the ALFA multi beam instrument.

The Abell cluster 1367, due to its proximity (z$\sim$0.0216, V$\sim$6500 \kms) and to 
the fact that it is currently forming at the intersection of two filaments in the Great Wall \citep{COGA04}, 
represents an ideal target for HI surveys (e.g. \citealp{sullivan81,chinca83,gav87,solanes01,peppo06}). It is ideal 
for the study of the properties of HI selected galaxies and 
environmental effects.
For these reasons, Abell 1367 is one of the region observed as a part of the 
Arecibo Galaxy Environment Survey (AGES, \citealp{auld06}), one 
of the new HI blind survey carried out with the ALFA multi-beam system\footnote{The other three main extragalactic ALFA surveys 
are ALFALFA \citep{alfaalfa05}, AUDS \citep{auds} and ZOA \citep{zoa}.}.
AGES aims to study the atomic hydrogen
properties of different galactic environments to faint sensitivity limits;
low HI masses (6$\times10^{8}~\rm M_{\odot}$, assuming a 200 \kms velocity width at 92.8 Mpc, the distance of the A1367 cluster) 
and column densities of $\sim$ 3$\times10^{18}~\rm cm^{2}$ (for a source that fills the beam). 
The environments AGES will survey range from apparent voids in the
large-scale structure of galaxies, via isolated spiral galaxies and their
haloes, to galaxy-rich regions associated with galaxy clusters and
filamentary structures.
AGES plans to cover $5\times4$ square degrees centered on Abell 1367.
Here we report on the early results obtained from observations of a subset of the whole 
AGES-A1367 region: a $5\times1$ square degree area centered on the cluster core.

\section{AGES observations and Data Reduction}
As part of the AGES, we observed the Abell 1367 region in May 2006 and April 2007 using 
the ALFA feed array at the 305-m Arecibo Telescope. 
A total of $\sim$63 hours observing time were allocated with an average observing time of $\sim$2.3 hours per night.
Since the allocated time was insufficient to cover all the AGES-A1367 region (5$\times$4 square degrees) 
we decided to give the highest priority to the cluster center and its immediate outskirts covering a strip 
of $\sim$5$\times$1 square degrees (11:34:00$<$R.A.(J.2000)$<$11:54:15, 19:15$<$Dec(J.2000)$<$20:20) 
centered on the cluster core. 
The observation and data reduction techniques adopted by AGES are extensively described in \cite{auld06}; here 
we will only provide a brief summary of them.\\
AGES uses the 7-feed ALFA multi-beam receiver and a spectral line backend capable of instantaneously 
recording spectra from the two linear polarizations of each beam and covering a bandwidth of 100 MHz.
The angular resolution is given by the size of the ALFA beams ($\sim$3.3$\times$3.8 arcmin) and the 
velocity resolution is $\sim$25kHz, corresponding to $\sim$ 5.5 \kms. 
Observations are performed in \emph{drift scan mode} (e.g. \citealp{alfaalfa05}): the array is kept
at a fixed azimuth and elevation while the sky drifts overhead. 
The Earth's rotation rate governs the on source time in a drift
scan. For Arecibo this means that each point in the sky takes $\sim$12 s
to cross the beam. Twenty-five separate scans are then required
to reach the 300 s/beam integration time. 
In order to compensate for the change in parallactic angle, and thus
achieve uniform sky coverage, ALFA must be rotated before every
scan and, in order to attain
fully sampled sky coverage, it is necessary to stagger the declination
(Dec.) of individual scans by $\sim$1 arcmin (half the beam separation).
We record data every second and total power values for each of the seven beams, both
polarizations and 4096 channels, are recorded as four-bit floating
point numbers. 
Calibration is performed at the beginning of every scan 
using a high-temperature noise diode
that is injected into each beam for a duration of 1 s. 
A uniform final depth of $\sim$ 300 sec was achieved for all of the area observed. 

Data reduction was performed using the AIPS++ packages LIVEDATA
and GRIDZILLA \citep{barnes01}, developed by the Australian
Telescope National Facility (ATNF). LIVEDATA
performs bandpass estimation and removal, Doppler tracking and
calibrates the residual spectrum. 
GRIDZILLA is a gridding package that co-adds all the spectra using
a suitable algorithm, to produce 3D data cubes. 
The Abell 1367 5$\times$1 square degrees region was gridded using a median gridding 
technique into 1$\times$1 square arcmin pixels, each of which contains a 4096 channel spectrum, and was smoothed 
at a velocity resolution of 10 \kms.
The AGES observing strategy is highly successful at reducing sidelobe variations by 
making a Nyquist-sampled map with every ALFA beam individually. Thus, when the observations 
are median combined, the variations across
the beams are removed, leaving a circular beam with symmetrical sidelobes at a $\sim$5-10\% level \citep{minchin07}.
The output data cube has two spatial dimensions and the spectral dimension
can be chosen by the user to be frequency, wavelength or velocity.
The noise distribution is fairly gaussian with a standard deviation of 0.8439$\pm$0.0003 mJy, consistent with 
the expected value \citep{auld06} for minimal smoothing of the beam. 
The cube is publicly available and can be downloaded at the following 
link $http://www.naic.edu/\sim ages/public\_data.html$.

\subsection{Data contamination: Radio frequency interference, the Milky Way and 3C264.}
Radio frequency interference (RFI) represents a constant source of contamination in certain 
regions of the observed spectral window.
In particular two main RFI sources were present in all the scans: 
the San Juan FAA radar at 1350 MHz and its 3rd harmonic at 
1387.8 MHz.
In addition an intermittent source of RFI (GPS satellite L3) 
appeared only in some of the scans at approximately 1381 MHz.
The strong constant sources at 1350 MHz (V$\sim$14620$\pm$140 \kms) and 1387.8 MHz (V$\sim$7009$\pm$19 \kms)
 completely obscure a part of the cosmic volume: in particular 
 the 1387.8 MHz RFI lies approximately at the velocity of Abell 1367 
 making part of the cluster volume inaccessible. 
The intermittent GPS source at 1381 MHz (V$\sim$8500 \kms) reduces our sensitivity 
limit and could significantly influence the number of sources detected 
in a frequency range $\sim$ 300 \kms wide.

In addition to the RFI another two sources of contamination affect our data.
The first one is the Galaxy: LIVEDATA is not able to correctly 
subtract the strong emission from the Milky Way, resulting in a blindness of our survey 
in about the velocity range -50$<V<$+50 \kms.
Moreover the Abell 1367 region includes the 3C264 radio galaxy (R.A.=11:45:05.5, Dec=+19:36:23). 
This strong radio source has a 1.4GHz continuum flux  of $\sim$5 Jy, producing 
strong spectral standing waves and making a $\sim$4 arcmin radius region around the galaxy inaccessible. 

\subsection{Source Extraction}
In every blind survey a crucial step in the data reduction is represented by the 
source extraction.
The objective is to obtain a catalogue of confirmed sources 
to well defined selection limits: as the signal-to-noise ratio (S/N) decreases, the 
contamination by spurious noise signals quickly rises. 
Detection techniques need to be optimized to discriminate real from spurious signals as well as possible. 

Currently source extraction is carried out using three different methods:
the cube is visually inspected independently by two of us (in this case RRA and LC) and by an automatic 
extractor (Polyfind, written by R. F. Minchin) based on a peak flux threshold.
The automatic extractor is essentially based on the cross-correlation with templates method described by \cite{davies01}. 
The program initially looks for peak values above a pre-defined value (in this case 4$\sigma$) and then 
cross-correlates with templates accepting the best fit as long as the correlation coefficient 
is above 0.75 and the total signal-to-noise ratio of the detection is above 4 \citep{davies04}.
Automatically extracted candidates are then checked 'by eye', visually inspecting the full AGES spectrum summed from 
a 5$\times5 \rm~ arcmin^{2}$ box around the position of the detection. 
The aim of this process is to quickly remove obvious spurious detections while not rejecting any real source.
These methods provided a list of 104 candidate HI sources\footnote{Here we do not include two extended 
High Velocity Cloud complexes extending over a great part of the observed region at 140$<V<$ 210  \kms. These 
systems are described in the Appendix of this paper.}, of which 78 were detected by all the three different signal extraction techniques.
In the following we will refer to these 78 objects as "sure" sources and to the remaining 26 as "possible" 
sources. 

\begin{figure}
\centering
\includegraphics[width=10cm]{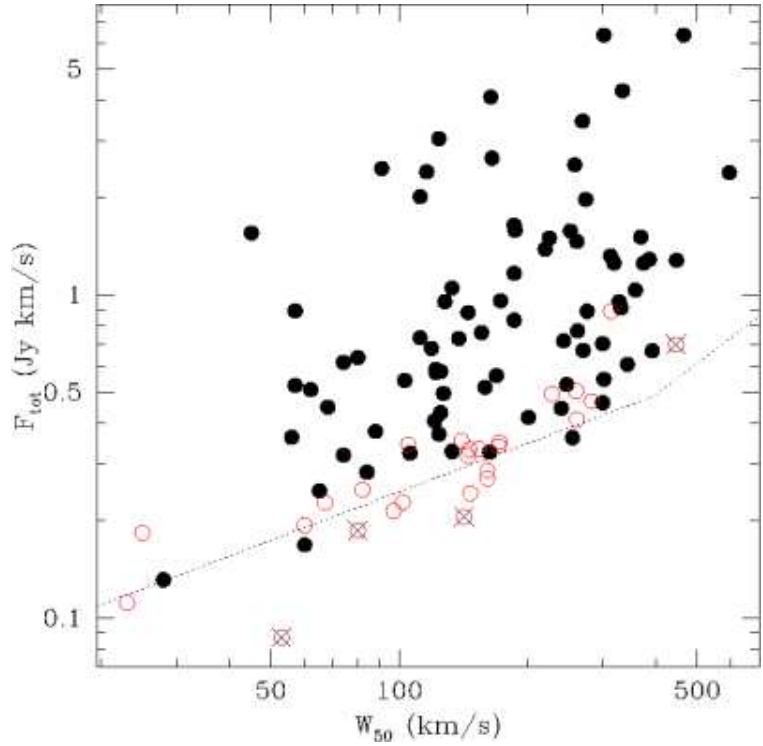}
\caption{The distribution of the 104 HI source candidates detected in the AGES-A1367 cube. 
Filled circles indicate "sure" sources (i.e. detected by all the three extraction methods adopted), 
empty circles are sources not detected by all the three extraction methods and investigated with 
follow-up Lwide observations (see Sec.\ref{secLw}). Crosses indicate those sources that 
have not been confirmed by follow-up observations. The dotted line shows the ALFALFA reliability limit, $S/N_{tot}=6.5$ \citep{saintonge07}.}
\label{FtotW}
\end{figure}

The distribution of the total 21 cm flux as a function of the line width at the 50\% level of peak maximum for our 104 candidate 
sources is shown in Fig.\ref{FtotW}.
Filled circles indicate "sure" sources while empty circles are "possible" sources.
The void in the bottom right part of the plot (i.e. large velocity widths and small fluxes) is due to a selection 
effect present in the source extraction technique: at fixed low total flux it is easier to detect 
a narrow HI source, since the survey is not flux but surface brightness limited.
In order to compare the AGES S/N threshold with the recent results presented by ALFALFA \citep{giovanelli07} we plotted in Fig.\ref{FtotW} a S/N limit of $S/N_{tot}$=6.5 for our survey, where \citep{saintonge07}:  
\begin{equation}
S/N_{tot} = \frac{1000\times F_{tot}}{W_{50}}\times\frac{w^{1/2}}{rms}
\label{eqSN}
\end{equation}
$F_{tot}$ is the total flux in Jy \kms, $W_{50}$ is the velocity width measured at the 50\% level, 
$w$ is either $W_{50}/(2\times~\Delta V)$ for $W_{50}<400$ \kms or 400/$(2\times~\Delta V)$ for $W_{50}\geq400$ \kms, 
$\Delta V$ is the spectral velocity resolution and $rms$ is the r.m.s. noise across the spectrum measured in mJy at 
$res.$ spectral resolution. In our case $res.$=10 \kms.
The great majority of our "sure" sources lie well above $S/N_{tot}$=6.5 while the "possible" sources 
have $S/N_{tot}\approx$6.5, in agreement with \cite{saintonge07}.
This suggests that also for our sample $S/N_{tot}$=6.5 is a reliable threshold for discriminating between 
"sure" and "possible" sources.

\section{L-Wide Radio Follow-up Observations}
\label{secLw}
In order to test the reliability of our extraction technique and the real nature of the 26 "possible" sources,  
10 hours of telescope time where scheduled with the Arecibo single pixel L-band wide (LBW) receiver in April 2007. 
The LBW observations were made covering a spectral band of 12.5 MHz (2048 spectral channels) centered at the expected source recessional velocity. 
The integration time varied between 5 and 10 minutes depending on the intensity of the source, reaching an 
rms in the range $\sim$0.5-0.8 mJy beam$^{-1}$ at a velocity resolution of 10 km s$^{-1}$.
All observations were taken using the position-switching technique, with each blank sky (or OFF) position observed 
for the same duration, and over the same portion of the telescope dish (Az and El) as the on-source (ON) observation. 
Each 5 min + 5 min ON + OFF pair was followed by a 10 s ON + OFF observation of a calibrated noise diode. 
None of the 26 observed sources lies within the sidelobes of a "sure" HI detection, making unlikely a sidelobe 
contamination of the L-Wide follow-up observations.
Of the 26 "possible" sources, only 4 (indicated with crosses in Fig.\ref{FtotW}) were not confirmed by Arecibo follow-up observations.
We therefore have a sample of 100 sure HI sources detected in the 5$\times$1 square degrees observed by AGES in the 
Abell 1367 region.   

\section{Source Catalogues}
All the 100 sources in the sample were inspected and parametrized  using the MIRIAD task MBSPECT \citep{miriad}.
The source position was estimated from a Gaussian fit to the moment map within the galaxy velocity range and 
the spectrum was then extracted from the data cubes. 
This spectrum is a weighted average, 
with the weight depending on the distance of each pixel from the fitted position. 
Velocity widths were 
then measured at the 20\% and 50\% levels ($W_{20}$ and $W_{50}$) relative to the peak signal, using 
a width-maximizing algorithm \citep{lewis83}. 
The uncertainties in each derived quantity are computed following the recipes 
given by \cite{kori04}.
In Table \ref{suretab} we present the main parameters of all the 100 HI sources detected in the cube, namely:\\
Col.1: HI source ID\\
Col.2-3: HI R.A. (J.2000) coordinate and relative error\\ 
Col.4-5: HI Dec. (J.2000) coordinate and relative error\\ 
Col.6: Heliocentric velocity and relative error ($cz$, i.e. optical reference frame), measured as the midpoint between the velocities 
where the flux density reaches 50\% of peak maximum level.\\
Col.7: Velocity width and relative error of the source line profile measured at the 50\% level of peak maximum.\\ 
The velocity widths are not corrected for instrumental broadening, turbulent motions, disk inclination 
or cosmological effects.\\
Col.8: Velocity width and relative error of the source line profile measured at the 20\% level.\\
Col.9: Peak Flux density and relative error in mJy.\\
Col.10: Total Flux and relative error in Jy \kms.\\
Col.11: Object flag, defined as follows: Flag 0 indicates sources detected by all the three independent 
methods used for source extraction making them reliable sources. Flag 1 indicates objects confirmed 
only after Lwide follow-up observations.
Flag 2 indicates objects which are contaminated by RFI: they are in general sure detections 
but the HI parameters (e.g. flux, velocity width, etc.) can be strongly affected and should not be used.\\
The AGES source catalogue and the spectra for all the HI sources are publicly available 
at the following link $http://www.naic.edu/\sim ages/public\_data.html$.

\begin{figure*}
\includegraphics[width=5cm]{vel.epsi}
\includegraphics[width=5cm]{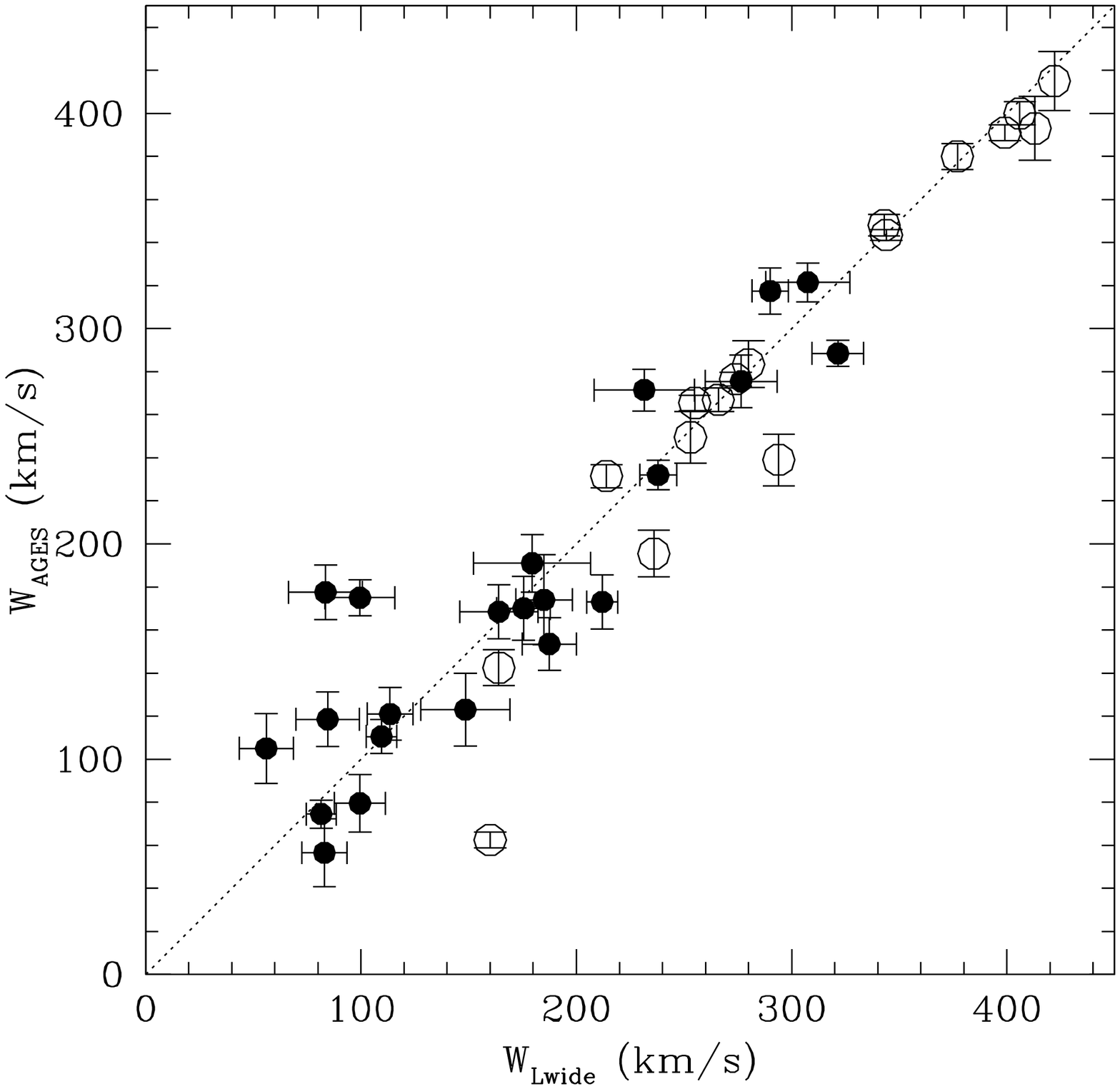}
\includegraphics[width=5.cm]{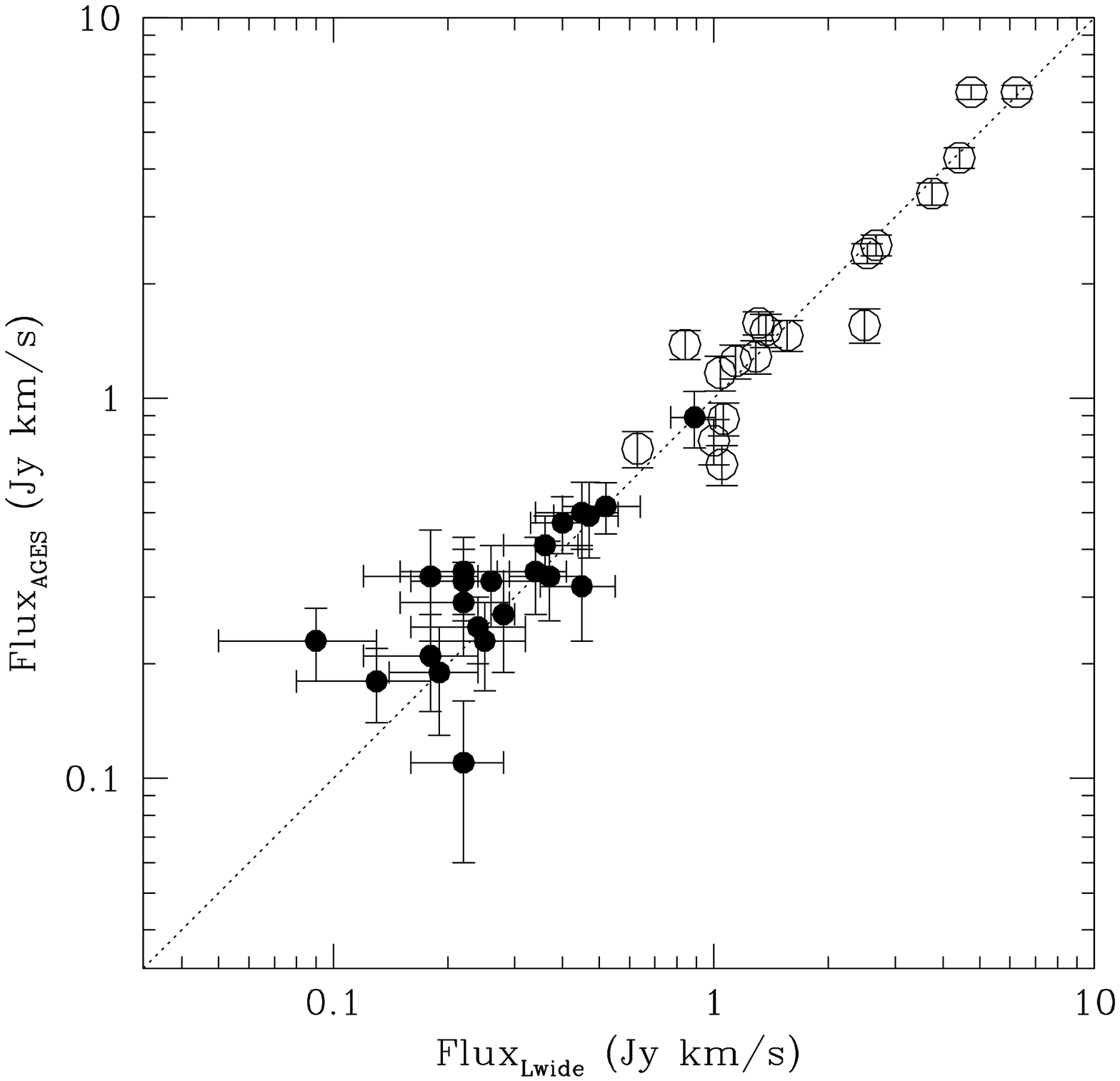}
\caption{Comparison of the recessional velocities (Left), the velocity widths (Center) and the HI fluxes measured by AGES and L-wide. Filled circles indicate the 22 sources confirmed during follow-up observations, empty circles 
show the 18 galaxies already known from the literature . The dotted lines indicate a one to one correlation.}
\label{lwide}
\end{figure*}

The results of the LBW follow-up observations are presented in Table \ref{lwidetab}, which lists the following:\\ 
Col.1: HI source ID\\
Col.2: Heliocentric velocity and relative error of the source ($cz$), measured as the midpoint between the channels 
at which the flux density drops to 50\%.\\
Col.3: Velocity width and relative error of the source line profile measured at the 50\% level.\\ 
The velocity widths are not corrected for instrumental broadening, turbulent motions, disk inclination 
or cosmological effects.\\
Col.4: Velocity width and relative error of the source line profile measured at the 20\% level.\\
Col.5: Total Flux and relative error in Jy \kms.\\
Col.6: The rms noise per channel at 10 km s$^{-1}$ velocity resolution.\\
In Fig.\ref{lwide} we compare the measurements obtained using the ALFA and L-wide receivers for the 22 sources confirmed during 
follow-up observations (filled circles). Although these are the lowest S/N sources within our sample, the agreement between the dataset is quite 
good with an average dispersion of 19 \kms in recessional velocity, 35 \kms in velocity width and $\sim$14\% 
in total flux.
    
\subsection{Comparison with previous works}
The Coma-A1367 supercluster is one of the regions most intensively investigated by optically 
selected HI surveys (i.e. \citealp{haynes97,peppo06}). However only 21 out of the 100 sources detected by AGES 
had a previous HI measurement.  
The high number of new HI detections in AGES ($\sim$79\%) confirms once more the importance 
of HI blind surveys for having an unbiased view of the neutral hydrogen distribution in the local Universe.
Among the 21 sources previously known 
\citep{gav89,haynes97,springob05,peppo06}, three of these objects (namely CGCG97-079, CGCG97-087 and CGCG127-052) are strongly contaminated by RFIs in the AGES cube, and cannot be used for a comparison with the literature. Our measurements 
for the remaining 18 objects are in satisfactory agreement with past works as illustrated in Fig.\ref{lwide} (empty circles). There is quite a good agreement in both fluxes and velocity width measurements and the average scatter is $\sim$10\% 
in flux and $\sim$ 28 \kms in velocity width.
In one case (CGCG97-138) our velocity width estimate differs considerably from literature values: \cite{gav89} 
gives a velocity width of 160\kms whereas we measure only 62\kms).
However our new L-wide measurement is consistent with the value obtained from the AGES cube (W=58 \kms). 
    
\subsection{Optical Counterparts}
The Abell 1367 region has been covered in the optical by the Sloan Digital Sky Survey (SDSS) as part of Data 
Release 5. We have used this homogenous dataset to look for optical counterparts of our HI sources. 
We cross-correlated our list of HI detections with the SDSS galaxy catalogue using a search radius 
of 4 arcmin. In the few cases in which no candidate optical counterpart was found, we visually inspected 
the SDSS plates to look for low surface brightness galaxies not included in the SDSS catalogue.
Once a candidate optical counterpart was found, we looked into the literature and NED 
for an estimate of its optical recessional velocity. In addition for 5 of the 46 candidates without 
an available redshift in the literature we obtained new optical recessional velocity estimates in January-Febrary 
2007 using the imaging spectrograph BFOSC attached to the Cassini 1.5 m telescope at Loiano (Italy). Observations and data reduction 
techniques applied are described in \cite{redshift}.

Confirmed optical counterparts are defined as those optical galaxies lying within 4 arcmin and having a recessional 
velocity within $\pm$200 \kms from the radio velocity. 
Candidate optical counterparts are galaxies which lie within 4 arcmin but do not have a published optical recessional velocity.
Of the 100 HI sources in our sample, 55 have confirmed optical counterparts, 
34 have a unique candidate (i.e. no optical recessional velocity available) optical counterpart, 
7 have two or more candidate counterparts and 4 sources appear to have no optical counterparts. 
Three of these objects are associated with the interacting groups in the outskirts of Abell 1367 described in the Appendix of this paper.
A list of the optical counterparts (confirmed and candidate) associated with the HI sources is presented in Table \ref{optcounttab}.

\begin{figure}
\centering
\includegraphics[width=8cm]{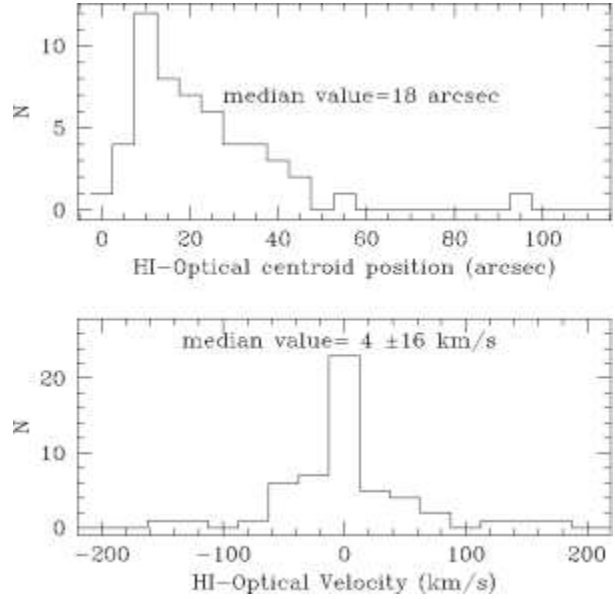}
\caption{Distribution of the positional (upper panel) and recessional velocity (lower panel) 
difference between the HI sources and their optical counterparts.}
\label{centroid}
\end{figure}

In Fig.\ref{centroid} we show the distribution of the difference between the optical and HI positions and recessional velocities 
for the 55 sources having confirmed optical counterparts. Optical and radio measurements show a very good agreement 
with a median offset of only 18 arcsec in position (consistent with the typical error of 16 arcsec in the estimate of the centroid 
for radio sources) and of $\sim$4$\pm$16 \kms in recessional velocity.
 
For all the HI sources having 1 optical counterpart (confirmed or candidate) we downloaded SDSS-DR5 $ugriz$ images 
\citep{SDSSDR5} and performed aperture photometry and surface brightness profile decomposition using the task {\it ellipse} within IRAF. 
The radial profiles have been constructed by integrating the available
images within elliptical, concentric annuli. The ellipticity and
position angles have been determined and then fixed using the 
$i$ band image. Total magnitudes, effective radii ($r_{e}$) and effective surface brightness ($\mu_{e}$, i.e. average surface brightness within $r_{e}$) have been determined 
following the procedure of \cite{gav00}.

\subsection{Comparison with other HI-blind surveys}
\begin{figure*}
\includegraphics[width=18cm]{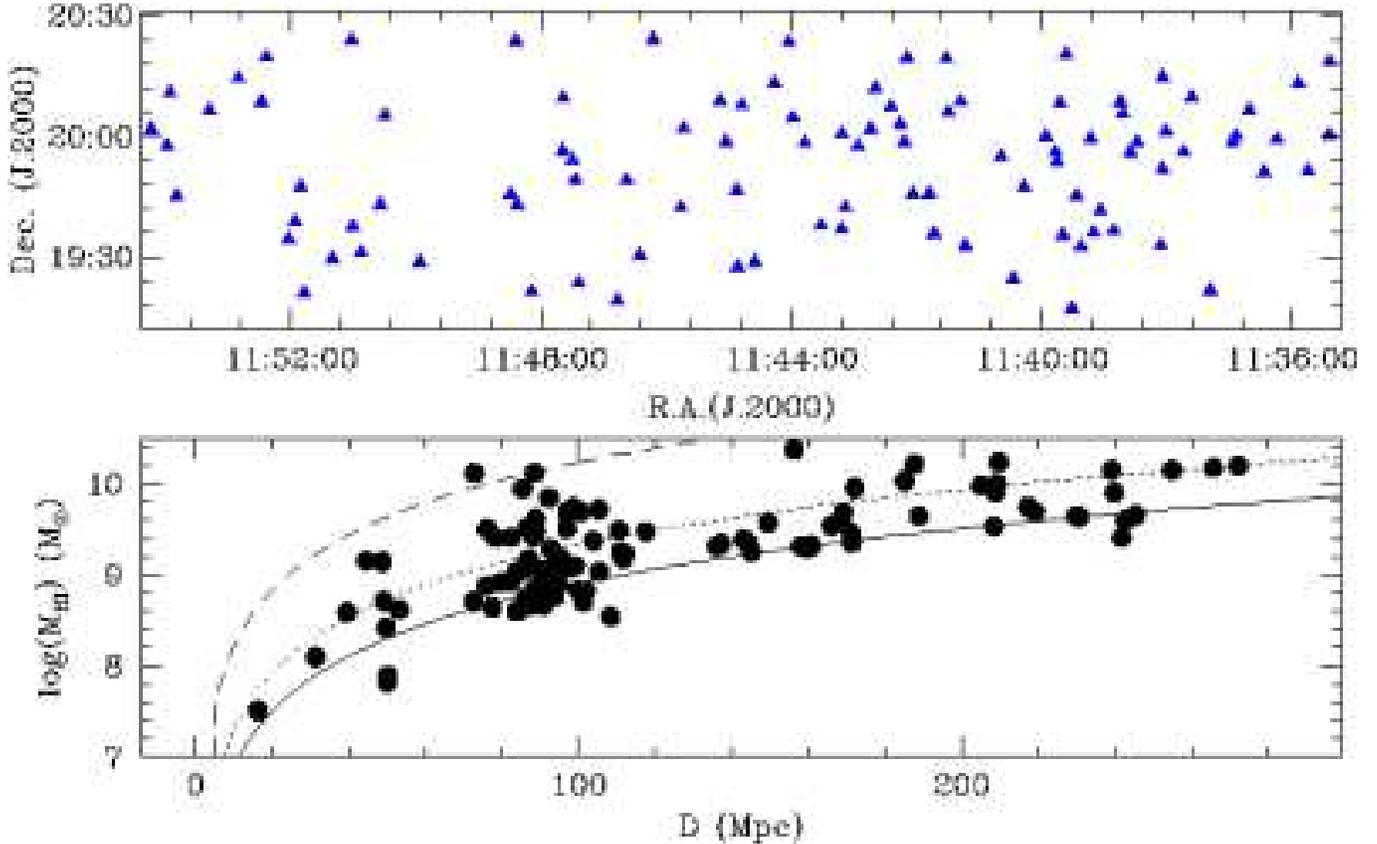}
\caption{The sky (upper panel) and HI mass vs distance (lower panel) distribution for 
the 100 HI sources in our sample. In the lower panel distances are obtained from the recessional velocities assuming 
that all galaxies are in the Hubble flow. The solid line shows the sensitivity limit for a $S/N_{tot}\sim6.5$ source having 
$W_{50}$ of 200\kms. The dotted line show the ALFALFA reliability limit for the same 
velocity width \citep{giovanelli07}. The dashed line indicates a flux integral of 7.3 Jy \kms, which 
corresponds  to the HIPASS 6.5$\sigma$ limit for a velocity width of 200\kms.}
\label{distrall}
\end{figure*}

Fig.\ref{distrall} illustrates the sky (upper panel) and the HI mass vs distance distributions 
(lower panel) for all the 100 sources detected in the A1367 cube. Over half of the sources (54) 
belongs to the Abell 1367 cluster and its outskirts, lying in the velocity range 4000$<V<$9000 \kms. 
This roughly explains why in this cube we detect twice the number of sources than in other AGES fields: excluding 
the Abell 1367 region (4000$<V<$9000) we are left with 46 sources consistent with the number found in 
NGC628 \citep{auld06}, NGC1156 (Auld et al. in prep.) and NGC7332 \citep{n7332}. 
The bottom panel of Fig.\ref{distrall} allows  a direct comparison of AGES  with ALFALFA \citep{alfaalfa05} showing that, in the same area of sky, ALFALFA would miss $\sim$ half of the sources 
detected here. This is quite expected since ALFALFA has an integration time $\sim$ 6.3 times shorter ($\sim$48 sec instead of $\sim$300 sec) than AGES. 
More impressive is the difference between AGES and the northern HIPASS 
extension \citep{hipassnorth}: whereas AGES finds 100 sources, HIPASS does not detect a 
single galaxy in the 5$\times$1 square degrees here studied.

\section{Physical Properties of the AGES-A1367 sample}
We use the HI data from AGES and the optical data from SDSS to derive 
the physical properties of our sample.
The HI mass ($M(HI)$) is defined as:
\begin{equation}
M(HI) = 2.36 \times 10^{5} D^{2} F_{tot}  ~~~~~~~~~(M_{\odot})
\end{equation} 
where $F_{tot}$ is the integrated HI flux in Jy \kms and D is the galaxy distance in Mpc.
For those galaxies belonging to Abell 1367 (4000$<V<$9000 \kms) we assume $D$=92.8 Mpc \citep{sakai00} 
while we assume $D=V/H_{0}$ with $H_{0}=$ 70 \kms $\rm Mpc^{-1}$ for the rest of the sample.
The stellar mass is obtained from the optical colors and luminosity as defined in \citep{bell03}:
\begin{equation}
log(M_{star}) = -0.222 + 0.864(g-r) + \frac{M(i) - 4.56}{-2.5}  ~~~~~~~~~(M_{\odot})
\end{equation} 
where $M(i)$ is the absolute magnitude in the $i$ band.
Therefore the total amount of baryons is $M_{bar}$=$M_{star}$+$M(HI)$ (e.g. \citealp{bell03b}).
This quantity does not include the contribution of helium and metals (usually assumed 
equal to 1.4$\times$M(HI)), the molecular hydrogen and any warm or hot gas component.
Therefore $M_{bar}$ must be considered as a lower limit of the real value.
Finally we define the galaxy dynamical mass as:
\begin{equation}
M_{dyn} = \gamma \frac{R_{75} \times W_{c}^{2}}{2G}
\end{equation}
where $G$ is the gravitational constant, $R_{75}$ is the radius containing 75\% of the $i$ band flux, 
$W_{c}$ is the HI velocity width (here we use the average between $W_{50}$ and $W_{20}$ HI velocity widths) corrected for inclination and $\gamma$ is 
the ratio between $R_{75}$ and the extension of the HI emission (i.e. $\gamma=R_{HI}/R_{75}$). $\gamma$ should lie 
between 2-3 \citep{salpeter96,swaters02}, but for our galaxies radio interferometric data (necessary to estimate $R_{HI}$) is not available, so this remains unknown and we leave it as a free parameter.  
The galaxy inclination is obtained from the ellipticity determined in the $i$ band via:
\begin{equation}
cos(i)^{2} = \frac{(1-e)^{2}-q_{0}^{2}}{1-q_{0}^{2}}
\end{equation}
where $e$ is the galaxy ellipticity and $q_{0}$ is the intrinsic axial ratio of the galaxy here assumed equal to 0.13 \citep{gio97}.  
We excluded from this analysis face-on galaxies having $e<0.1$ (20 galaxies), for which it is not possible 
to accurately estimate the dynamical mass. 

\begin{figure*}
\includegraphics[width=18.5cm]{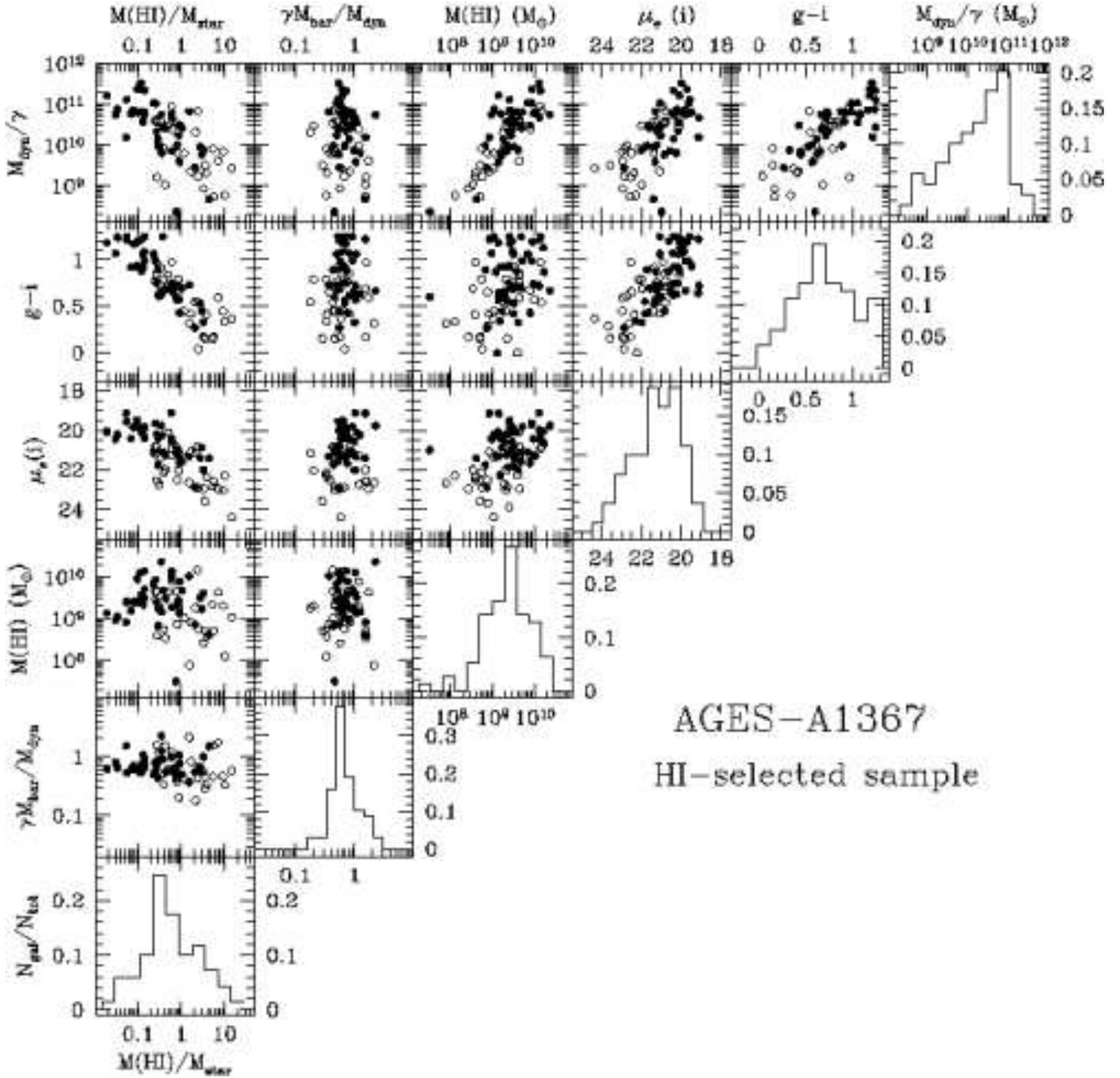}
\caption{Distributions and correlations between HI and optical properties for the 69 galaxies in our sample having 1 optical counterpart and $e>0.1$. 
From left to right the ratio of the gas to stellar mass, the baryons fraction, the total HI mass, the effective 
surface brightness in $i$ band, the $g-i$ color and the dynamical mass are presented.
Filled dots are objects with confirmed optical counterparts while empty dots indicates 
objects with optical counterpart candidates. }
\label{paramall}
\end{figure*}

\begin{figure}
\centering
\includegraphics[width=8cm]{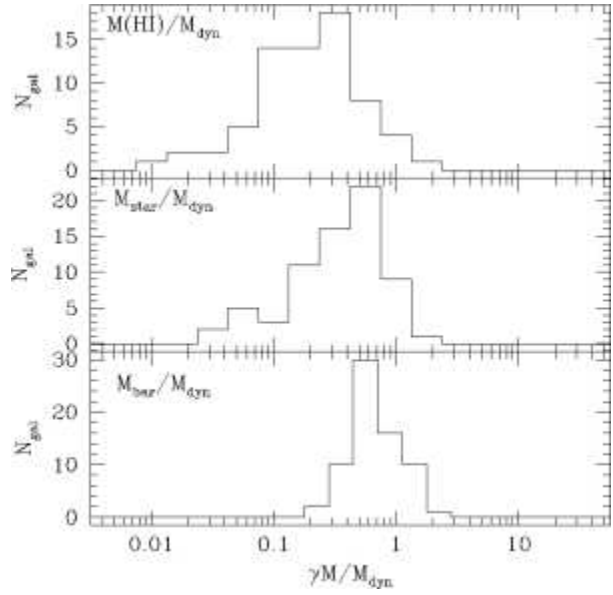}
\caption{The distribution of the gas to dynamical mass (top), star to dynamical mass (middle) and 
baryon to dynamical mass (bottom) ratios for our sample.}
\label{massratios}
\end{figure}

The correlations between different optical and HI quantities and their distributions for our sample
are presented in Fig.\ref{paramall}.
From left to right: the ratio of the gas to stellar mass, the baryon fraction ($M_{bar}/M_{dyn}$), the total HI mass, the effective 
surface brightness in $i$ band, the $g-i$ color and the dynamical mass are presented.
Only galaxies with one optical counterpart and $e<0.1$ (69 galaxies) are shown: filled dots are objects with confirmed 
optical counterparts while empty dots indicate objects with optical counterpart candidates. 
Our sample covers almost three orders of magnitude in both 
HI and dynamical mass. The optical surface brightness distribution shows 
a significant tail at low surface brightness 
($\mu(i)_{e}\sim$22.5-24 mag $\rm arcsec^{-2}$) as usually observed in HI
selected sample (e.g. \citealp{spitzak98}). HI selected samples are often considered as composed of only 
blue, gas-rich galaxies. This is not the case for our sample which 
spans over 1 magnitude in the $g-i$ color, as is typically observed in optically 
selected samples \citep{baldry04}, and covers a wide range of gas to stars ratios
($M_{gas}/M_{star}$): from gas poor objects ($M_{gas}/M_{star}\sim$ 0.1) to extremely gas rich galaxies ($M_{gas}/M_{star}\sim$ 10). 
More massive galaxies appear to be redder, have higher surface brightness and lower gas content than 
dwarf systems. This is also found to be the case in optically selected samples \citep{phenomen}.
Well known relations like color-mass, surface brightness-mass and hydrogen-dynamical mass are here rediscovered.
Unfortunately the small number of objects in our sample does not allow a more detailed comparison between 
the slopes, scatters and distributions of the relations followed by HI and optically selected galaxies.
We postpone this kind of analysis to the moment when we will have a significantly large AGES database at our disposition.\\ 
The most intriguing result presented in Fig.\ref{paramall} is the absence of any correlation involving the baryon fraction.
On the contrary all galaxies in our sample appear to have almost the same baryon fraction $<log(M_{bar}/M_{dyn})>$=(-0.2-log($\gamma$))$\pm$0.13 independently from their mass, surface brightness or gas content.
The little dispersion in the baryon fraction distribution is in fact well within the observational errors. 
Assuming 0.1 dex error in the HI mass (see Sec.4.1), in the stellar mass \citep{bell03} and 
in the dynamical mass, the uncertainty on the estimate of the baryon fraction is $\sim$0.17 dex.
The gas and stellar mass fraction cover a larger dynamical range (i.e. have a larger dispersion: $<log(M_{star}/M_{dyn})>$=(-0.4-log($\gamma$))$\pm$0.24 and $<log(M(HI)/M_{dyn})>$=(-0.6-log($\gamma$))$\pm$0.3)) than the baryon fraction, as shown in Fig.\ref{massratios} and accordingly  to the Kolmogorov-Smirnov test the baryon fraction distribution is not compatible 
with the gas and star fraction distributions  at a 99.99\% confidence level. 

We can therefore conclude that our sample shows approximately the same baryon fraction.
A constant baryonic mass fraction in galaxies implies a direct correlation 
between dynamical mass and baryonic mass: exactly what is usually required for the 
Tully-Fisher relation \citep{mcgaugh98,mcgaugh00}.  
This result is also consistent with the recent study of an optically selected sample 
of extremely low mass dwarf galaxies carried out by \cite{geha06}.
To our knowledge, 
this is the first confirmation of a constant baryonic fraction coming from an HI selected sample.
This observational evidence is also consistent with models of galaxy formation \citep{bullok01,crain07} but 
in contradiction with recent models of galaxy evolution which 
invoke supernova feedback and other physical mechanisms to remove baryons 
preferentially from low mass systems and to reproduce the observed scaling relations \citep{cole00,governato07}. 
Even if our sample includes low surface brightness galaxies 
it appears that physical mechanisms that preferentially remove baryons from low mass galaxies 
are not effective in the mass range and selection criteria of our sample.
We remark that the only way to obtain a correlation between the baryon fraction and the dynamical 
mass in our sample would be to 
assume that $\gamma$ (i.e. the ratio of the HI to optical radius) is a strong function of the 
mass and/or the contribution 
of molecular hydrogen, hot and warm gas to the total baryon mass varies with the mass of the galaxy.
We cannot a priori exclude the first possibility, even if unlikely since \cite{cayatte94} found 
that the ratio of the HI to optical radius varies by only 20\% along 
the Hubble sequence. Conversely, the molecular hydrogen represents 
only $\sim$15\% of the total gas reservoir in normal, late-type galaxies \citep{boselligdust} and 
the contribution of hot and warm gas components is  a few percents ($\leq$1-2\%) of the total gas mass (e.g. \citealp{tschoske01,higdon06})
making the second scenario extremely unlikely.

\section{Comparing the optical and HI view of the Abell cluster 1367}
More than half of the HI sources detected in the cube belong to the Abell 1367 volume.
This includes not only the cluster core but also a part of the Great Wall: a 
large scale filament of galaxies connecting Abell 1367 to the Coma cluster \citep{zabludof93}.
All galaxies in the filament lie approximately at the same distance from 
us as Abell 1367, implying that the system 
A1367+Great Wall constitutes a volume limited sample. 
At the distance of Abell 1367 ($\sim$92.8 Mpc) the surveyed sky area 
has a physical size of 8.1$\times$1.6 $\rm Mpc^{-2}$ and our sensitivity limit
for a source with $W_{50}$=200 \kms  is $\sim$6$\times10^{8}~\rm M_{\odot}$.
Since the virial radius of Abell 1367 is $\sim$ 2.3 Mpc  \citep{GIRARD98}, our region covers the cluster virialized region and its immediate neighborhood.  

In order to compare the HI and optical properties of the Abell 1367 region 
we extracted an optically selected sample of galaxies from SDSS-DR5.
We select all galaxies in the AGES region having $g<17$ mag (corresponding to $L_{g}>2\times10^{9}~\rm L_{\odot}$ at the Abell 1367 distance): 259 galaxies in total, of which 208 have optical 
redshift available and 155 are confirmed cluster members (4000$<V<$9000 \kms).

We note that the time necessary to carry out a HI targeted survey of this optically selected sample (reaching 
the same AGES noise level) would be similar to the observing time needed by AGES to cover all the 
5$\times$1 square degrees in Abell 1367.

\subsection{Sky distribution}
\begin{figure*}
\centering
\includegraphics[width=18cm]{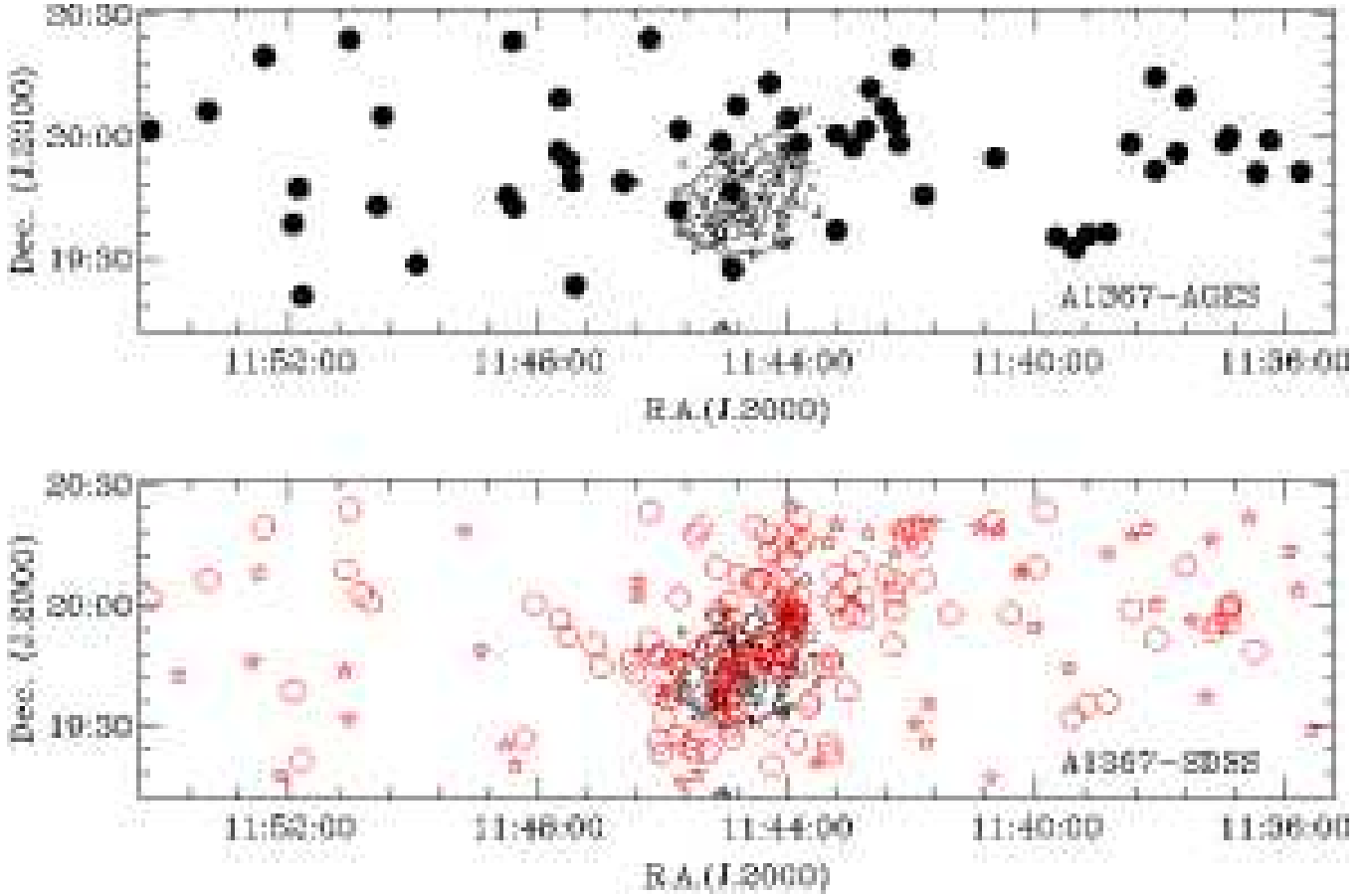}
\caption{
Upper panel: Sky distribution of the members (4000$<V<$10000 \kms) of the Abell 1367 cluster in the HI selected 
sample. Lower panel: Sky distribution of galaxies in the SDSS optically selected sample ($g<$17 mag).
Empty circles indicate confirmed cluster members and empty stars galaxies without redshift available.
The black contours indicate the X-ray emission from A1367 as measured by ROSAT.}
\label{skya1367}
\end{figure*}

The sky distribution of the 54 HI sources belonging to the A1367 region (4000$<V<$9000 \kms) is shown 
in Fig.\ref{skya1367}. It is interesting to note that no significant over-density of sources is observed corresponding 
to the cluster center: the only HI galaxy lying well within the X-ray cluster 
contours belongs to a small group of galaxies infalling for the first time into 
the cluster, suggesting that it is just entering the cluster and it is only projected 
on the center of Abell 1367  \citep{big}.
This appears much more evident when we compare the sky distribution of our sample with the typical sky distribution 
obtained for the optically selected sample. In Fig.\ref{skya1367} we show the distribution of 
the 155 confirmed cluster members (empty circles) and of the 51 galaxies without redshift available\footnote{
The majority of galaxies without redshift have a magnitude $g<16.5$ mag, suggesting that a not negligible 
fraction is composed by background galaxies \citep{COGA04}.}  (empty stars) extracted from the optically selected sample.

\begin{figure}
\centering
\includegraphics[width=8.5cm]{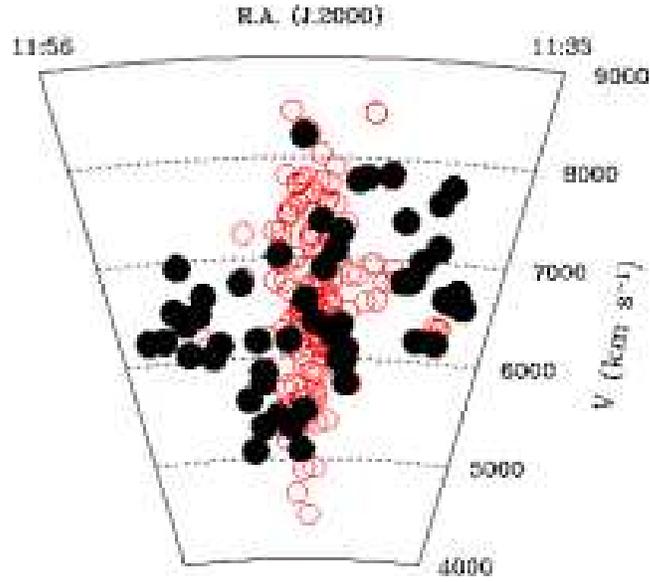}
\caption{The wedge diagram of the AGES-A1367 region in the velocity range 4000$<V<$9000 \kms. 
Filled circles indicate the HI detections while empty circles are galaxies belonging to the SDSS optical 
selected sample ($g<$17 mag).}
\label{wedge}
\end{figure}

The optical selected sample is strongly clustered on the cluster center and almost half of the 
sources detected lie within the X-ray emitting region.
Moreover the "Finger of God" feature, typical of clusters of galaxies, is not observed in the HI selected sample (see Fig.\ref{wedge}).
Conversely the HI-selected galaxies show a different pattern in the wedge diagram shown in Fig.\ref{wedge}: galaxies 
on the east side of the cluster have a lower recessional velocity than galaxies in the west side. 
This probably suggests the presence of two different infalling directions 
into the cluster core as proposed by \cite{COGA04}.

The differences observed between the optically and HI selected samples are somehow expected since 
clusters of galaxies are made HI deficient structures through the combination of the morphology-density relation \citep{dressler80,whitmore} and of the "gas-density" relation \citep{giova85,peppo06}. This is also consistent 
with recent determinations of the correlation function for HI selected samples \citep{meyer07,basilakos07} which show that HI galaxies are among the least clustered objects in the Universe.

\begin{figure}
\centering
\includegraphics[width=8.5cm]{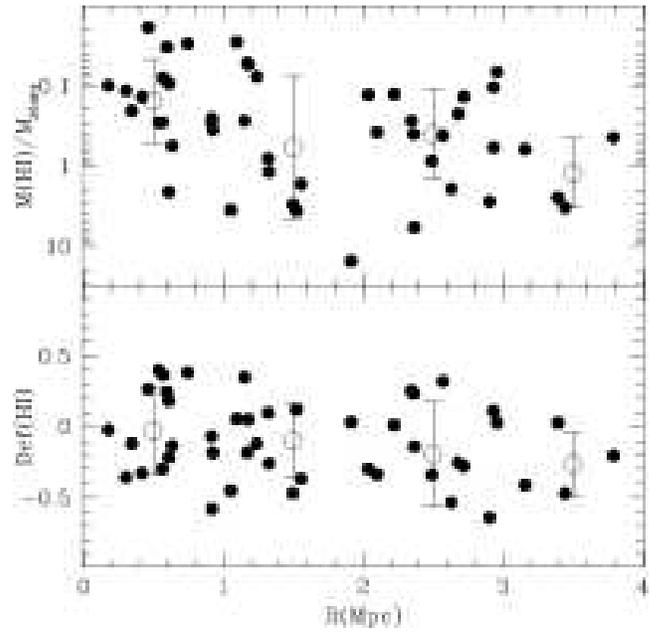}
\caption{The distribution of the gas to star mass ratio (upper panel) and of the HI deficiency as a function 
of the projected distance from the center of Abell 1367. Empty circles show the average values and standard 
deviations obtained in bins 1 Mpc wide.}
\label{defradius}
\end{figure}

The influence of the harsh cluster environment is barely visible when we
look for radial variations of the gas to star mass ratio and of the HI-deficiency (see Fig.\ref{defradius}).
The HI deficiency is defined as the difference, in logarithmic units, between the 
observed HI mass and the value expected from an isolated galaxy 
with the same morphological type T and optical linear diameter D:
HI DEF = $<\log M_{HI}(T^{obs},D^{obs}_{opt})> - log M^{obs}_{HI}$ \citep{haynes}.
We used the equations in \cite{solanes96} to calculate the expected HI mass from the optical diameter.
The HI deficiency does not significant vary with the cluster-centric distance (Fig.\ref{defradius}, bottom panel), contrary 
to what usually observed in optically selected samples (i.e. \citealp{haynes,solanes01,peppo06}).
A slight decrease on the average value of the gas to star mass ratio is observed in correspondence of the cluster center, however also in this case no strong gradients have been found.
We can therefore conclude that when observed in HI the Abell cluster 1367 almost completely disappears and the only evidence of its hidden presence is a slightly higher number of galaxies with a low gas to star mass ratio near the cluster core.

\subsection{The HI mass distribution and optical luminosity function.}
The Abell 1367 local volume can be used for a more detailed analysis of selection effects representing a unique dataset for which complete samples independently selected 
at optical and radio wavelengths are available.
We are particularly interested in the comparison of the HI mass distribution and optical luminosity function estimated 
from the two samples. In the past, different authors have used optically selected samples 
to determine the HI mass distributions of galaxies in the local Universe obtaining 
results significantly different from radio selected samples \citep{briggs93,spring05hi,gavazzi05}.
We therefore used the optically selected sample extracted from SDSS-DR5, complete to $g<17$ mag, in order 
to compare the HI distribution and the $g$ band luminosity function for optically and HI selected galaxies.

Contrary to the HI selected sample, the SDSS sample is not complete in redshift, but the 
redshift completeness rapidly drops below 50\% for $g\sim16.5$  mag ($\sim3 \times 10^{9}~\rm L_{\odot}$).
This complicates the estimate of the luminosity function at low luminosities.
In order to overcome the redshift incompleteness of cluster samples we used the \emph{completeness corrected} method proposed by \cite{depropris03}. This method is based on the assumption that the spectroscopic sample (i.e. galaxies with redshift available) is 'representative' of the entire cluster, i.e. the fraction of galaxies that are cluster members is the same in the (incomplete) spectroscopic sample as in the (complete) photometric one.

\begin{figure}
\centering
\includegraphics[width=8.5cm]{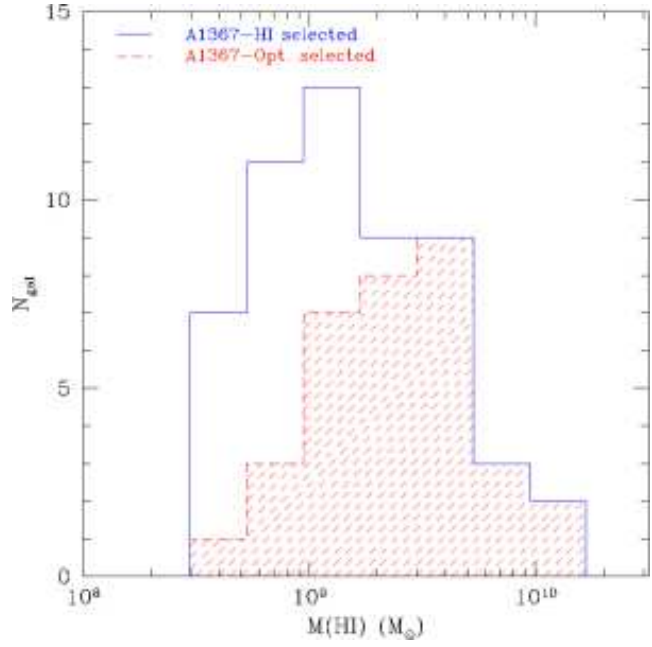}
\caption{The HI mass distribution for galaxies in the A1367 volume. 
The solid histogram indicates the HI selected sample and the shaded histogram the optically selected sample extracted from SDSS-DR5 ($g<17$ mag).}
\label{LFHI}
\end{figure}

The HI mass distributions obtained from the optically and HI selected samples are shown in Fig.\ref{LFHI}.
As expected the two samples produce HI distributions significantly different.
The optical selected sample drops quicker at low HI masses and does not include nearly half of the HI sources 
in the A1367 volume. The two samples start to differ at $M(HI)\sim$2$\times10^{9}~\rm M_{\odot}$, well before our sensitivity  limit.  Low luminosity, low surface brightness, gas rich objects are in fact not included in the optical sample, but they represent a significant fraction of the total HI budget in Abell 1367.
This result is consistent with the recent work by \cite{spring05hi} who showed that the faint end slope of the HI mass function for an optically selected sample is less steep than the one obtained from a 21 cm selected sample.
Our analysis suggests that this difference is mainly due to selection effects and 
that HI distributions obtained from optically selected samples cannot be generally considered as good proxies 
of the HI mass function in the local Universe.     

\begin{figure}
\centering
\includegraphics[width=8cm]{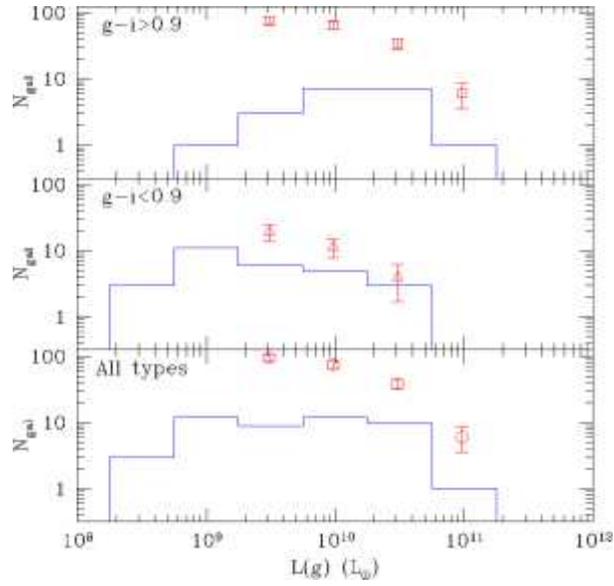}
\caption{Each panel shows the $g$ band luminosity function for galaxies in the A1367 volume as obtained 
from the optically (empty circles) and HI (histogram) selected sample.
All galaxies in the two samples are shown in the bottom panel, while blue ($g-i<$0.9) and red galaxies 
($g-i>$0.9) are shown in the middle and upper panel respectively.}
\label{LFopt}
\end{figure}

The HI and optically selected samples also differ in the estimate of the optical luminosity functions (Fig.\ref{LFopt}, lower panel).
The luminosity function (LF) for the HI selected galaxies is flatter and includes fewer objects (i.e. lower $\Phi^{*}$) than the 
one obtained from the optically selected sample. 
A great part of this difference is due to the overabundance of red ($g-i>$0.9), gas poor galaxies in the center of Abell 1367: more than  80\% of bright red galaxies ($L>10^{9}~\rm L_{\odot}$) in the A1367 region are 
not detected at 21 cm (see Fig.\ref{LFopt}, upper panel).
The difference between the two samples is less evident if we compare the luminosity functions 
obtained only for blue galaxies (Fig.\ref{LFopt}, middle panel), but even in this case the HI sample 
misses $\sim$30\% of blue cluster galaxies.

\begin{figure}
\centering
\includegraphics[width=8cm]{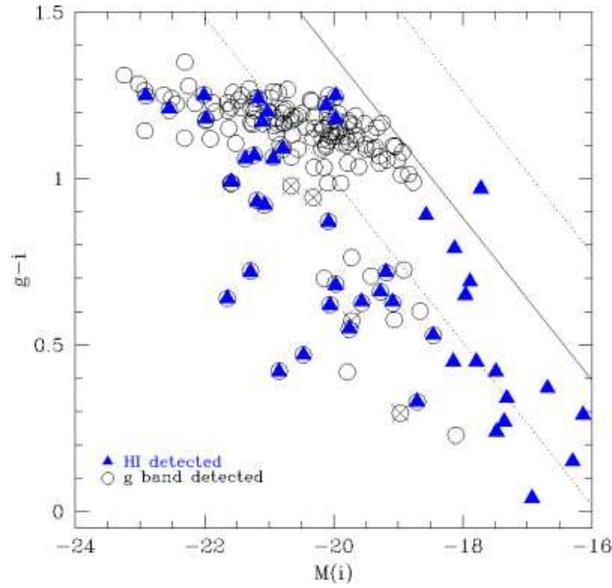}
\caption{The $g-i$ color $M(i)$ magnitude relation for confirmed cluster members in the Abell 1367 region. Triangles indicate galaxies detected in the AGES cube. Crosses indicate those galaxies lying in 
the RFIs velocity range or within the beam occupied by brighter sources. The solid (dashed) line indicate our
sensitivity limit ($\pm$1 $\sigma$) obtained converting the HI sensitivity into color limit using the $g-i$ versus $M(HI)/L(i)$ relation given by Eq.\ref{ml}. 
}
\label{colmag}
\end{figure}

This result is quite interesting since we generally associate HI deficiency to red colors. 
In order to further investigate the properties of these blue galaxies not detected in HI, in Fig.\ref{colmag} we compare the color $g-i$, $i$ magnitude relation for  the confirmed cluster members in the optically (empty circles) and HI selected (triangles) samples. 
Almost all the HI galaxies lie on the blue sequence, but the opposite is not true: not all the galaxies in the blue sequence are detected in HI. 
This could be due to a number of different reasons. 
First of all we have to take into account our sensitivity limit ($M(HI)\sim$6$\times10^{8}~\rm M_{\odot}$ for W$_{50}$ $\sim$ 200 \kms) which we need to relate to optical color and magnitude. To do so we have used the correlation between the HI mass and $i$ band luminosity ratio and the $g-i$ color observed for our sample.
The best bisector linear fitting gives the following relation:
\begin{equation}
\label{ml}
g-i = (-0.61\pm0.05) \times \log\big(\frac{M(HI)}{L(i)}\big) + (0.57\pm0.04)
\end{equation}
with a dispersion of 0.37 dex. We can use this equation to estimate our sensitivity limit, which is indicated by the solid line in Fig.\ref{colmag}.  We consider undetected blue galaxies those objects not detected by AGES, lying outside the one $\sigma$ region and with a $g-i$ color bluer than 0.9 mag: 10 galaxies in total.
Some of these sources lie within (in both space and velocity) strong HI galaxies and can be missed, due to the size of the Arecibo beam (FWHM$\sim$3.5 arcmin) and to the fact that in these cases the HI emission is totally assigned to the bright source. Moreover some galaxies have a recessional velocity within the RFIs frequencies and cannot be detected. 
Excluding these cases (indicated with crosses in Fig.\ref{colmag}, Left)
we are left with eight blue galaxies which are not detected in HI. 
These objects have a $g$ band luminosity in the range 2$\times$10$^{9} < L_{g}<$ 10$^{10}$ L$_{\odot}$ (i.e. -19.4$<M(g)<$-17.9 mag), implying that we are dealing with intermediate mass galaxies and not dwarf systems. 
Moreover their broad-band morphology and structural parameters are indistinguishable from the other blue HI detected 
galaxies.
The limits obtained from Eq.\ref{ml} imply that the undetected galaxies have a HI deficiency $\geq$0.37, suggesting 
that we are dealing with blue, deficient objects, consistent with the marginal detection of two of these galaxies (CGCG97-092, CGCG97-093) by deeper surveys \citep{peppo06} implying an HI deficiency 0.31 and 0.57 respectively. 
This suggests that we are dealing with the same population of HI poor, star forming spirals 
originally observed in nearby clusters by \cite{kennicutt84}.
Only when observed in H$\alpha$\footnote{The H$\alpha$ emission of a galaxy is due to the hydrogen ionized in HII regions by massive ($>$10 M$_{\odot}$), young ($<$ 2 $\times$ 10$^{7}$ yr) stars. } these galaxies appear different from 
healthy spirals, showing 
a truncation of the star forming disk.
In particular, in all but one object for which H$\alpha$ data are available (8 galaxies, \citealp{jorge02}) the H$\alpha$ radius (e.g. the isophotal radius within $10^{-16.5}~\rm  ergs~cm^{-2}~s^{-1}~\AA^{-1}~arcsec^{-2}$) is  $\sim$2-4 times less extended than the $r$ band 24th mag $\rm arcsec^{-2}$ isophotal radius (see Table \ref{nodettab}).
This is consistent with various independent studies \citep{cati05,koop06,review} which have 
shown that the HI deficiency is associated with a truncation of the star forming disk.
All the deficient objects lie at a projected distance of less than $\sim$1 degree from the center of Abell 1367, suggesting 
that the absence of HI and the truncation of the star formation disk is likely due to their interaction with the cluster environment. The fact that these objects have lost almost 40\% of their original gas content but their broadband morphology and optical 
colours are still those typical of normal star forming spirals implies that they are recent arrivals within the cluster 
environment. It appears that the neutral hydrogen has been already stripped well within the optical radius quenching the star formation in the galaxy outskirts, but A type stars have not yet died and the galaxies still appear as blue as they were before losing their gas.  Given that H$\alpha$ traces stars younger than $2\times10^{7}$ yr and the typical lifetime of A stars is $\sim$1 Gyr, we can speculate that our galaxies started to loose their gas $\sim$100Myr ago, in agreement with the typical timescale for gas stripping observed in Virgo cluster galaxies (e.g. \citealp{vollmer01,vollmer04,n4569,cortese07}) and predicted by numerical models (e.g. \citealp{ABAM99,shioya02,roediger07}). 
In less than 1 Gyr, when all the A stars are dead, these objects will lie in the $g-i$ red sequence and will not be any  
different from the bulk of the gas-poor (red) galaxy cluster population.
Only future detailed spectroscopic investigation will allow us the better estimate the time scale of this transformation. 
Finally, it is interesting to note that this population of blue gas-poor galaxies 
corresponds to $\sim$ one third of the whole population of blue galaxies in the A1367-volume, 
supporting the idea that A1367 is a cluster still young and a considerable number of objects are 
infalling into the cluster core for the first time \citep{COGA04}. 

\section{Summary \& Conclusions}
We have presented the results of a HI blind survey of 5$\times$1 square degrees centered on the core of Abell 1367 
as part of the Arecibo Galaxy Environment Survey.
One hundred HI galaxies have been detected (79 new measurements and 50 new redshifts), half of which belong to the Abell 1367 volume.
Comparing the properties of  the HI selected galaxies with an optically selected sample extracted from SDSS-DR5 we have shown that 
the large scale distribution of galaxies is strongly wavelength dependent: in HI the cluster core almost completely disappears 
and HI selected galaxies are homogeneously distributed in the cluster volume.
The difference between the two samples is even more evident when we compare their luminosity and HI distribution.
Our optically selected sample misses a considerable number of low luminosity, low surface brightness galaxies underestimating 
by almost a factor 2 the number of HI rich galaxies present in the cluster volume. On the contrary our HI sample 
contains a factor 3 less bright galaxies than the optical one. These are mainly ellipticals or early type, red spirals but 
also a not negligible fraction of the blue sequence cluster spirals is missed.
All these blue gas-poor objects are located near the edge of the X-ray cluster contours suggesting that their gas has been recently 
($\sim$ 100 Myr) stripped by ram pressure so that the galaxies are already gas poor whereas their stellar populations are still young.\\
Although the spatial and number distribution of samples selected at different wavelengths are significantly different, the internal properties of galaxies seem not to be significantly wavelengths dependent. 
HI galaxies seem to follow the same scaling relations observed in optical and bigger galaxies are redder, have 
a higher surface brightness and a lower gas content than lower luminosity galaxies.
The most interesting result of our analysis is that the baryon fraction of our sample appears to be almost constant, suggesting the difference in the star to gas ratio observed between bright and faint galaxies is only due to a different star formation efficiency (i.e. the time scale necessary to convert gas into stars).
This is the first time that such a result is obtained from an HI selected sample 
and even if consistent with models of galaxy formation it suggests that 
during their evolutionary history less massive galaxies do not lose a larger fraction of their baryons 
than bigger objects.

The analysis presented in this paper points out the importance of HI blind surveys like 
AGES in order to have an unbiased view of the properties of galaxies in the local Universe. 
Once the survey is at a more advanced stage and the size of our sample is significantly 
bigger than the one presented here it will be possible to carry out more detailed quantitative studies and comparisons 
between optically and HI selected samples, gaining more insight into the evolutionary history of nearby galaxies.         

\section*{Acknowledgments}
LC wishes to thank all the AGES consortium and in particular Elias Brinks \& Noah Brosh for useful discussions and comments on this manuscript.
LC is supported by the UK Particle Physics and Astronomy Research Council. 

This work is based on observations collected at Arecibo Observatory.
The Arecibo Observatory is part of the National Astronomy and Ionosphere Center, which is operated by Cornell University under a cooperative agreement with the National Science Foundation.
We wish to thank the Arecibo Observatory for granting us the time to complete this part of the project and its staff 
for the help during the observations and data reduction.
The TAC of the Loiano telescope is acknowledged for the time allocation to this project. This research has made use of the NASA/IPAC Extragalactic Database, which is operated by the Jet Propulsion Laboratory, California Institute of Technology, under contract to NASA and of the GOLDMine database \citep{goldmine}.

    Funding for the SDSS and SDSS-II has been provided by the Alfred P. Sloan Foundation, the Participating Institutions, the National Science Foundation, the U.S. Department of Energy, the National Aeronautics and Space Administration, the Japanese Monbukagakusho, the Max Planck Society, and the Higher Education Funding Council for England. The SDSS Web Site is http://www.sdss.org/.

    The SDSS is managed by the Astrophysical Research Consortium for the Participating Institutions. The Participating Institutions are the American Museum of Natural History, Astrophysical Institute Potsdam, University of Basel, University of Cambridge, Case Western Reserve University, University of Chicago, Drexel University, Fermilab, the Institute for Advanced Study, the Japan Participation Group, Johns Hopkins University, the Joint Institute for Nuclear Astrophysics, the Kavli Institute for Particle Astrophysics and Cosmology, the Korean Scientist Group, the Chinese Academy of Sciences (LAMOST), Los Alamos National Laboratory, the Max-Planck-Institute for Astronomy (MPIA), the Max-Planck-Institute for Astrophysics (MPA), New Mexico State University, Ohio State University, University of Pittsburgh, University of Portsmouth, Princeton University, the United States Naval Observatory, and the University of Washington.

\section*{Appendix: Interesting objects}
\begin{figure*}
\includegraphics[width=8.8cm]{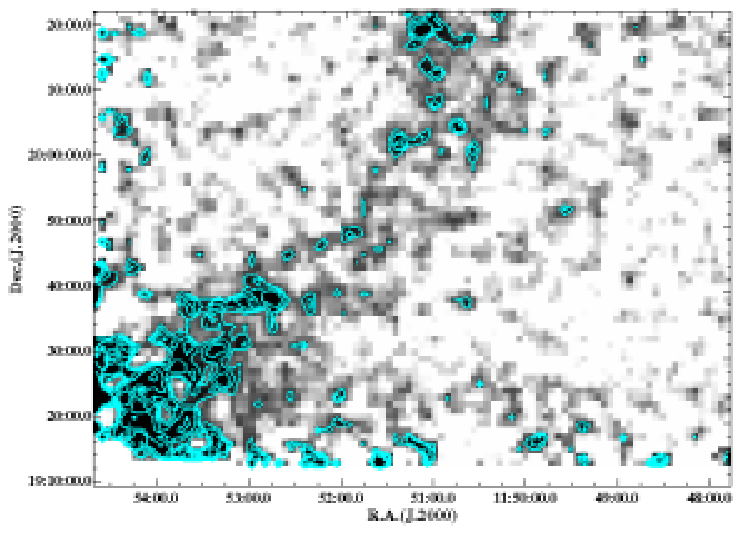}
\includegraphics[width=8.8cm]{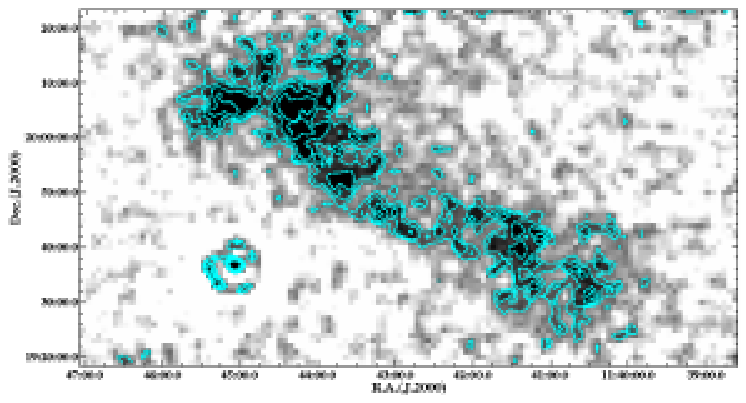}
\caption{The two High velocity clouds complexes detected in the AGES-A1367 cube. The HI column density maps 
were obtained integrating the line emission in the data cube within the velocity range 160$<V<$210 \kms (left) and 140$<V<$175 \kms (right). 
Contours are superposed to the moment map:  0.08, 0.1, 0.12, 0.16 Jy \kms $\rm beam^{-1}$.}
\label{HVC}
\end{figure*}

\begin{figure*}
\includegraphics[width=8.8cm]{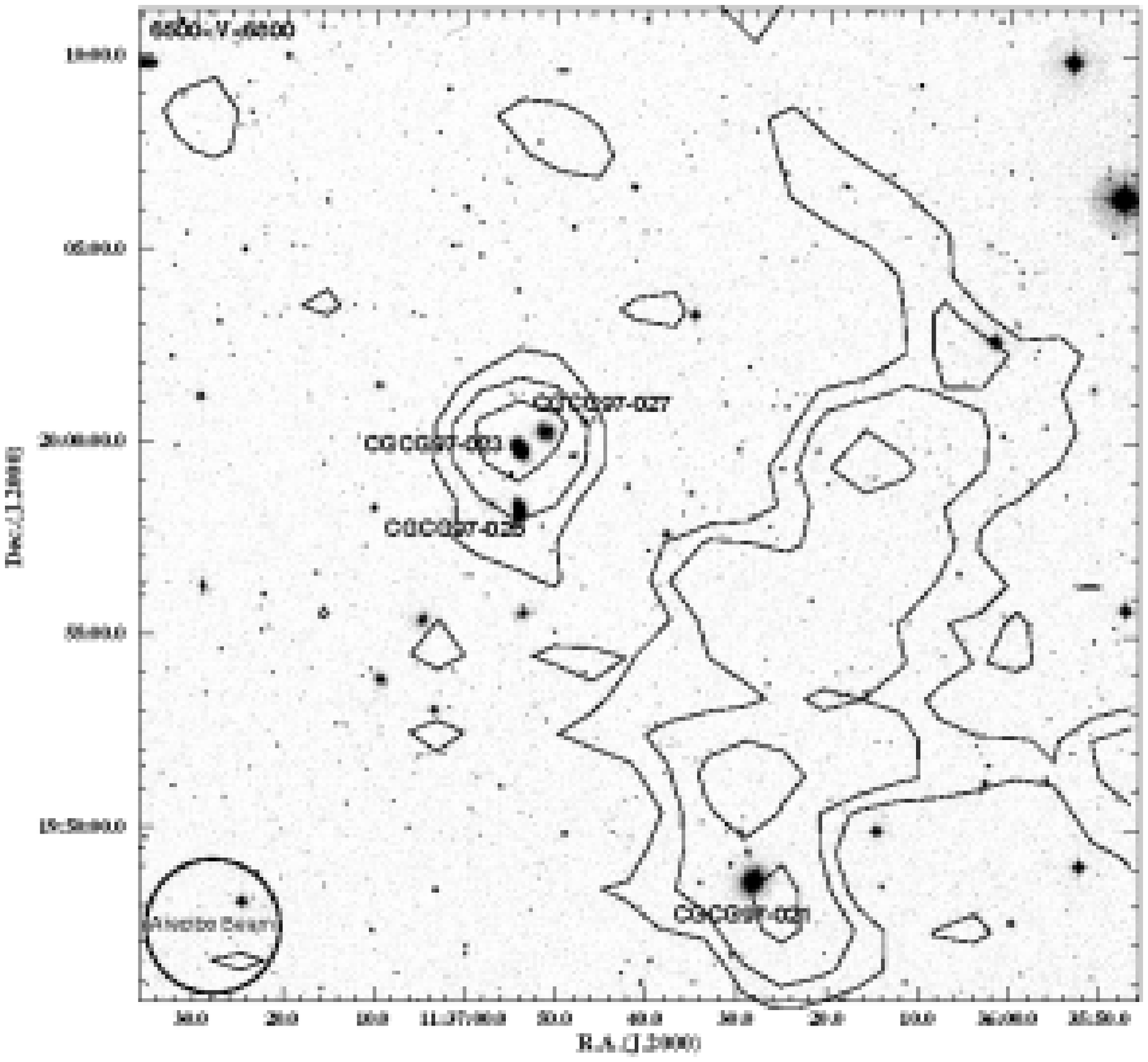}
\includegraphics[width=8.8cm]{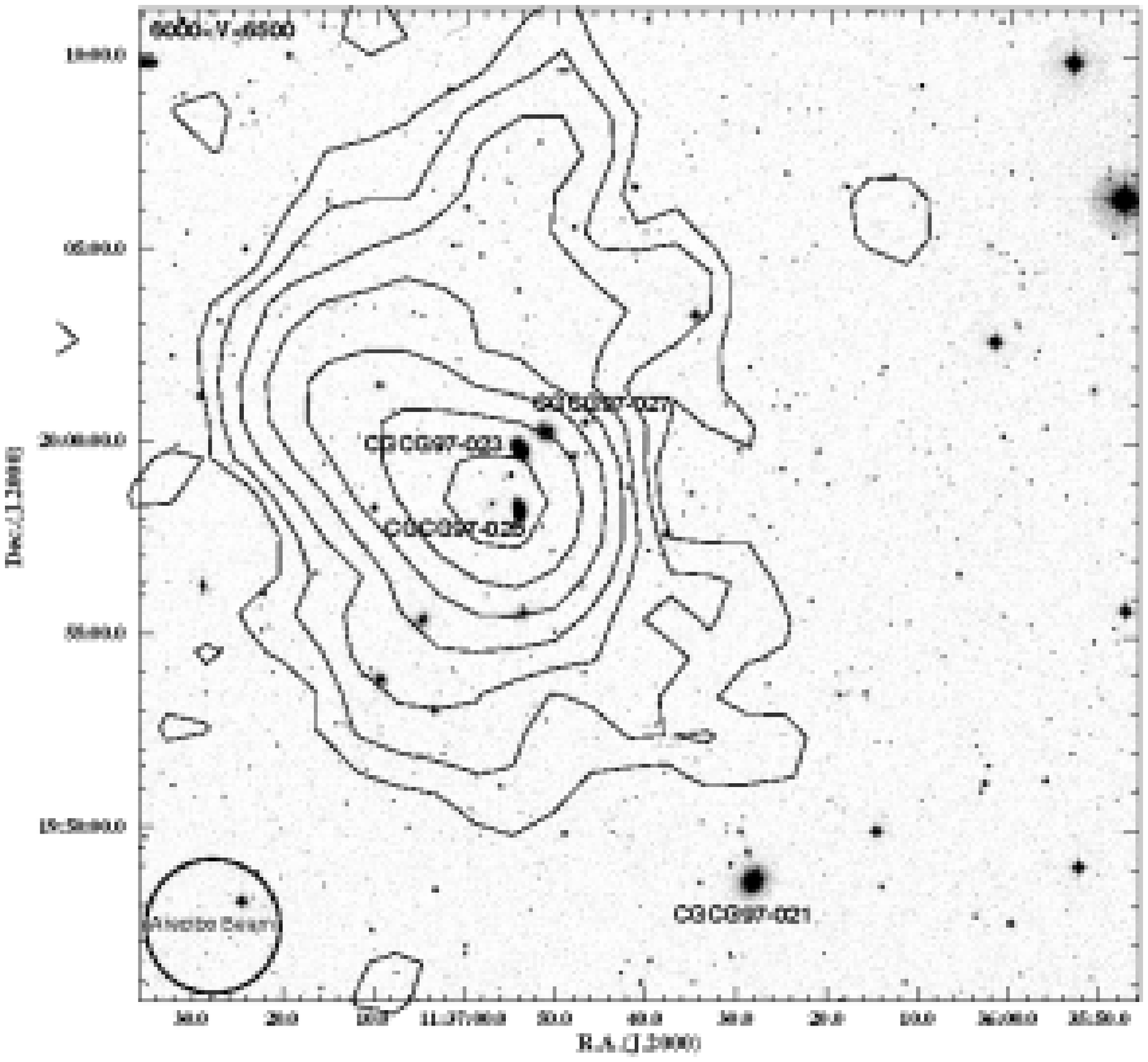}
\caption{The CGCG97-027 group as seen on the DSS-blue plates. HI columns density contours were obtained from moment maps of the AGES 
data cube in the velocity range 6500$<V<$6850 \kms (left) and 6000$<V<$6500 \kms are superposed. 
Contours levels are:  0.2, 0.3, 0.5 Jy \kms $\rm beam^{-1}$ (left) and  0.3, 0.46, 0.71, 1.1, 1.7, 2.6, 4 Jy \kms $\rm beam^{-1}$ (right). }
\label{97027}
\end{figure*}

\begin{figure}
\centering
\includegraphics[width=8.8cm]{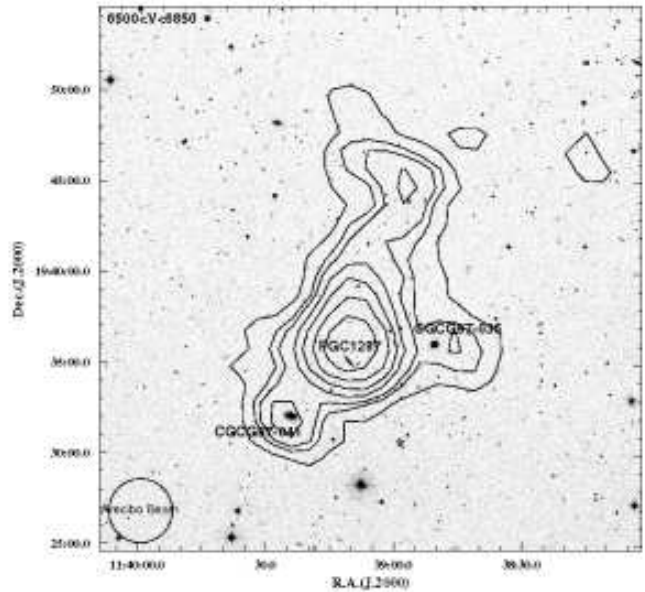}
\caption{The CGCG97-041 group as seen on the DSS-blue plates. Radio contours obtained from moment maps of the AGES 
cube in the velocity range 6500$<V<$6850 \kms. 
Contours levels are:  0.18, 0.26, 0.36, 0.52, 0.73, 1.05, 1.5 Jy \kms $\rm beam^{-1}$. }
\label{97041}
\end{figure}

\paragraph*{High Velocity Clouds}
Two extended features well separated from the Milky Way emission and probably associated with High Velocity Clouds (HVCs)
have been detected in the velocity range 140$<V<$210 \kms (Fig.\ref{HVC}).
The first lies in the range 11:50$<R.A.<$11:55 and probably extends outside the region observed. The brightest region 
is in the south-east corner of the cube (11:54:00, +19:20:00, 160$<V<$210 \kms) and a low column density stream seems to extend to the north-west.
The second and strongest HVC occupies great part of the east part of the cube and extends over 1.5 degrees (11:40$<R.A.<$11:46, 19:25$<Dec.<$20:15) with a velocity in the range 140$<V<$175 \kms. 
\paragraph*{CGCG97-027-group}
The CGCG97027-group is located at a projected distance of $\sim$2 
degrees ($\sim$3.2 Mpc) from the centre of Abell 1367. It is composed of 
two star forming spirals CGCG97027 (V$\sim$6630 \kms) and CGCG97026 (V$\sim$6220 \kms) and one 
lenticular CGCG97023 (V$\sim$6320 \kms). 
At a projected distance of $\sim$13 arcmin ($\sim$0.7 Mpc) SW from the group lies the elliptical galaxy CGCG97-021  (V$\sim$6648 \kms). Our data show the presence of diffuse HI 
emission all over the region occupied by the four bright Zwicky galaxies (see Fig.\ref{97027}). 
In particular and high velocity 6500$<V<$6800 \kms streams extends from CGCG97-021 to the north (Fig.\ref{97027}, left) and two of the HI sources without optical counterparts belong to this stream (AGES J113614+195910 and AGES J113626+195102).
Extended emission is also detected in the north part of the CGCG97-027 group in the velocity range 6000$<V<$6500 \kms (Fig.\ref{97027}, right). 
\paragraph*{CGCG97041-group}: The CGCG97041-group is located at a projected distance of ~1 
degrees (1.6 Mpc) from the centre of Abell 1367 and, as the CGCG97027-group, it is composed of 
two spirals CGCG97036 (V$\sim$6595 \kms) and AGC210559 (V$\sim$6825 \kms) and one 
lenticular CGCG97041 (V$\sim$6778 \kms). 
We detect HI emission extended over $\sim$200 kpc north of the three group members, apparently not associated with either of them (see Fig.\ref{97041}).

\paragraph*{AGES J113939+193524} This source ($V=$ 7382 \kms) has no optical counterpart catalogued in the SDSS database. 
However in the SDSS plates a low surface brightness ($\mu (i) \sim$ 24 $\rm mag~arcsec^{-2}$ galaxy is present 
in correspondence of the HI coordinates. We considered this object as candidate optical counterpart and 
labeled it as SDSSLSB in Table \ref{optcounttab}.

\paragraph*{AGES J114239+201150} This source lies at a projected distance of $\sim$5.8 arcmin from the starburst galaxy 
CGCG97-068. No galaxies or extended features are found in deep CFHT B-band \citep{goldmine} and 
GALEX NUV-FUV \citep{CORBGA05} data in correspondence with the peak of HI emission, suggesting that 
it could be a stream of gas stripped from CGCG97-068.
Unfortunately the distance of the source from CGCG97-068 is consistent with the region in which the first sidelobe 
is expected and the fact that AGES J114239+201150 lies at exactly the same recessional velocity of the starburst galaxy 
makes impossible to discard the possibility that this source is sidelobe contaminated.

VLA D-configuration observations has been recently obtained for the interacting groups 
described above and for CGCG97-068 and will be used to study in detail 
the properties of these sources. The results of these observations will be 
presented in a forthcoming work.

\paragraph*{AGES J114809+192109} This source  ($V=$ 11252 \kms) lies at the south edge of our cube and it is not clearly  
associated with any optical galaxy.

\clearpage
\onecolumn


\clearpage

\begin{table*}
\tiny
\caption{{HI parameters for the 100 sources detected in the Abell 1367 cube.}
\label{suretab}} 
\[
\begin{array}{ccccccccccc}
\hline
\noalign{\smallskip}
HI-ID  & R.A.  & \sigma_{R.A.} & Dec. & \sigma_{Dec.} & V & W50  & W20 & F_{peak} & F_{tot}   & flag \\
           & \rm (J.2000)  &  \rm sec. & \rm (J.2000) &  \rm arcsec             & \rm km\,s^{-1}    & \rm  km\,s^{-1}  & \rm  km\,s^{-1}    & \rm mJy    &  \rm Jy~  km\,s^{-1} &     \\
\noalign{\smallskip}
\hline
\noalign{\smallskip}
AGES~J113444+201217 & 11:34:44.4 & 0.7 & 20:12:17& 11 & 9602 \pm 6 & 127  \pm 12 & 181 \pm 18 & 5.5 \pm 0.7 & 0.50 \pm 0.08 & 0\\
AGES~J113508+201210 & 11:35:08.8 & 0.7 & 20:12:10& 10 & 16152 \pm 7 & 105 \pm 14 & 137 \pm 20 & 3.5 \pm 0.6 & 0.35 \pm 0.08 & 1\\
AGES~J113523+201823 & 11:35:23.8 & 0.7 & 20:18:23& 10 & 10464 \pm 6 & 300 \pm 11 & 344 \pm 17 & 5.0 \pm 0.7 & 0.71 \pm 0.10 & 0\\
AGES~J113524+200009 & 11:35:24.2 & 0.8 & 20:00:09& 13 & 9523 \pm 3 & 283  \pm 7 & 294 \pm 10  & 3.4 \pm 0.5 & 0.47 \pm 0.08 & 1\\
AGES~J113544+195120 & 11:35:44.8 & 0.8 & 19:51:20& 11 & 6564 \pm 7 & 84   \pm 14 & 120 \pm 20 & 3.7 \pm 0.6 & 0.28 \pm 0.07 & 0\\
AGES~J113553+201308 & 11:35:53.9 & 0.7 & 20:13:08& 11 & 1142 \pm 4 & 62   \pm 7 & 103 \pm 11  & 8.8 \pm 0.7 & 0.51 \pm 0.07 & 0\\
AGES~J113614+195910 & 11:36:14.3 & 0.8 & 19:59:10& 13 & 6713 \pm 11 & 119 \pm 22 & 273 \pm 33 & 4.9 \pm 0.6 & 0.68 \pm 0.10 & 0\\
AGES~J113626+195102 & 11:36:26.5 & 0.8 & 19:51:02& 12 & 6609 \pm 9 & 122  \pm 19 & 204 \pm 28 & 4.8 \pm 0.7 & 0.57 \pm 0.10 & 0\\
AGES~J113641+200622 & 11:36:41.2 & 0.7 & 20:06:22& 10 & 18604 \pm 8 & 276 \pm 17 & 336 \pm 25 & 4.6 \pm 0.7 & 0.89 \pm 0.13 & 0\\
AGES~J113653+200003 & 11:36:53.7 & 0.7 & 20:00:03& 11 & 6637 \pm 6 & 262  \pm 12 & 305 \pm 18 & 4.7 \pm 0.6 & 0.77 \pm 0.11 & 0\\
AGES~J113657+195837 & 11:36:57.1 & 0.7 & 19:58:37& 10 & 6190 \pm 3 & 302  \pm 7 & 394 \pm 10  & 26.6 \pm 1.4 & 6.38 \pm 0.28 & 0\\
AGES~J113718+192129 & 11:37:18.4 & 0.7 & 19:21:29& 10 & 14348 \pm 5 & 328 \pm 11 & 355 \pm 16 & 4.9 \pm 0.7 & 0.95 \pm 0.13 & 0\\
AGES~J113736+200934 & 11:37:36.0 & 0.7 & 20:09:34& 11 & 7739 \pm 3 & 369  \pm 7 & 391 \pm 10  & 8.7 \pm 0.9 & 1.51 \pm 0.15 & 0\\
AGES~J113743+195559 & 11:37:43.7 & 0.8 & 19:55:59& 13 & 7135 \pm 8 & 106  \pm 15 & 144 \pm 23 & 3.4 \pm 0.6 & 0.32 \pm 0.08 & 2\\
AGES~J113801+200106 & 11:38:01.0 & 0.8 & 20:01:06& 12 & 19052 \pm 5 & 314 \pm 10 & 329 \pm 15 & 4.5 \pm 0.8 & 0.90 \pm 0.15 & 1\\
AGES~J113804+195146 & 11:38:04.5 & 0.7 & 19:51:46& 10 & 6203 \pm 3 & 220  \pm 6 & 243 \pm 9   & 8.1 \pm 0.7 & 1.39 \pm 0.12 & 0\\
AGES~J113804+201441 & 11:38:04.5 & 1.0 & 20:14:41& 11 & 7580 \pm 4 & 60   \pm 7 & 80 \pm 11   & 3.7 \pm 0.4 & 0.17 \pm 0.04 & 0\\
AGES~J113805+193250 & 11:38:05.9 & 0.7 & 19:32:50& 10 & 17843 \pm 3 & 331 \pm 7 & 352 \pm 10  & 6.0 \pm 0.7 & 0.91 \pm 0.11 & 0\\
AGES~J113828+195823 & 11:38:28.5 & 0.7 & 19:58:23& 11 & 6989 \pm 2 & 133  \pm 4 & 144 \pm 6   & 7.9 \pm 0.7 & 0.33 \pm 0.06 & 2\\
AGES~J113835+195550 & 11:38:35.5 & 1.0 & 19:55:50& 17 & 13218 \pm 6 & 301 \pm 12 & 334 \pm 18 & 2.7 \pm 0.4 & 0.52 \pm 0.07 & 1\\
AGES~J113842+200527 & 11:38:42.4 & 0.8 & 20:05:27& 11 & 3513 \pm 4 & 28   \pm 7 & 49 \pm 11   & 4.7 \pm 0.6 & 0.13 \pm 0.04 & 0\\
AGES~J113844+200821 & 11:38:44.8 & 0.7 & 20:08:21& 10 & 3113 \pm 1 & 124  \pm 3 & 138 \pm 4   & 25.7 \pm 1.4 & 3.04 \pm 0.19 & 0\\
AGES~J113850+193622 & 11:38:50.8 & 1.8 & 19:36:22& 21 & 6811 \pm 6 & 122  \pm 11 & 196 \pm 17 & 6.1 \pm 0.6 & 0.59 \pm 0.07 & 0\\
AGES~J113904+194129 & 11:39:04.2 & 0.7 & 19:41:29& 11 & 16984 \pm 7 & 161 \pm 14 & 194 \pm 21 & 4.0 \pm 0.7 & 0.29 \pm 0.08 & 1\\
AGES~J113910+193558 & 11:39:10.3 & 0.7 & 19:35:58& 10 & 6783 \pm 7 & 116  \pm 13 & 383 \pm 20 & 15.4 \pm 0.9 & 2.40 \pm 0.15 & 0\\
AGES~J113913+195915 & 11:39:13.2 & 1.0 & 19:59:15& 12 & 10153 \pm 6 & 255 \pm 12 & 292 \pm 18 & 3.5 \pm 0.5 & 0.36 \pm 0.07 & 0\\
AGES~J113922+193232 & 11:39:22.6 & 0.8 & 19:32:32& 13 & 6783 \pm 3 & 252  \pm 5 & 282 \pm 8   & 9.8 \pm 0.7 & 1.58 \pm 0.11 & 0\\
AGES~J113926+194459 & 11:39:26.7 & 0.7 & 19:44:59& 15 & 17182 \pm 12 & 145\pm 23 & 203 \pm 35 & 2.7 \pm 0.6 & 0.32 \pm 0.09 & 1\\
AGES~J113930+191706 & 11:39:30.9 & 0.7 & 19:17:06& 10 & 3438 \pm 3 & 57   \pm 6 & 85 \pm 9    & 15.3 \pm 1.3 & 0.89 \pm 0.12 & 0\\
AGES~J113936+202020 & 11:39:36.9 & 0.7 & 20:20:20& 12 & 16743 \pm 4 & 358 \pm 7 & 370 \pm 11  & 5.2 \pm 0.8 & 1.03 \pm 0.15 & 0\\
AGES~J113939+193524 & 11:39:39.5 & 0.7 & 19:35:24& 10 & 7382 \pm 4 & 57   \pm 8 & 104 \pm 12  & 8.5 \pm 0.7 & 0.52 \pm 0.07 & 0\\
AGES~J113942+200817 & 11:39:42.4 & 0.8 & 20:08:17& 12 & 16920 \pm 9 & 25  \pm 18 & 185 \pm 27 & 5.3 \pm 0.6 & 0.18 \pm 0.04 & 1\\
AGES~J113945+195341 & 11:39:45.2 & 0.7 & 19:53:41& 10 & 11177 \pm 4 & 74  \pm 7 & 103 \pm 11  & 4.6 \pm 0.5 & 0.32 \pm 0.05 & 0\\
AGES~J113947+195559 & 11:39:47.3 & 0.7 & 19:55:59& 10 & 10933 \pm 3 & 164 \pm 6 & 246 \pm 10  & 24.8 \pm 1.3 & 4.09 \pm 0.21 & 0\\
AGES~J113956+195955 & 11:39:56.1 & 0.7 & 19:59:55& 11 & 10090 \pm 4 & 240 \pm 7 & 248 \pm 11  & 3.1 \pm 0.6 & 0.45 \pm 0.09 & 0\\
AGES~J114016+194715 & 11:40:16.9 & 0.7 & 19:47:15& 11 & 3476 \pm 3 & 68   \pm 6 & 96 \pm 9    & 6.9 \pm 0.6 & 0.45 \pm 0.06 & 0\\
AGES~J114027+192429 & 11:40:27.5 & 0.7 & 19:24:29& 10 & 3411 \pm 3 & 91   \pm 5 & 141 \pm 8   & 26.4 \pm 1.4 & 2.46 \pm 0.17 & 0\\
AGES~J114039+195455 & 11:40:39.0 & 0.7 & 19:54:55& 11 & 7839 \pm 3 & 186  \pm 6 & 214 \pm 10  & 6.7 \pm 0.6 & 0.83 \pm 0.09 & 0\\
AGES~J114113+193240 & 11:41:13.6 & 1.0 & 19:32:40& 14 & 10011 \pm 7 & 260 \pm 14 & 291 \pm 20 & 3.4 \pm 0.6 & 0.50 \pm 0.10 & 1\\
AGES~J114118+200829 & 11:41:18.0 & 0.7 & 20:08:29& 11 & 14616 \pm 4 & 156 \pm 7 & 186 \pm 11  & 5.8 \pm 0.6 & 0.76 \pm 0.08 & 0\\
AGES~J114129+200550 & 11:41:29.5 & 1.1 & 20:05:50& 21 & 14591 \pm 8 & 146 \pm 16 & 194 \pm 25 & 3.0 \pm 0.5 & 0.33 \pm 0.07 & 1\\
AGES~J114131+201912 & 11:41:31.5 & 0.9 & 20:19:12& 18 & 3504 \pm 9 & 23   \pm 18 & 90 \pm 26  & 4.6 \pm 0.7 & 0.11 \pm 0.05 & 1\\
AGES~J114143+193536 & 11:41:43.4 & 0.7 & 19:35:36& 10 & 16787 \pm 4 & 125 \pm 8 & 157 \pm 13  & 6.1 \pm 0.7 & 0.58 \pm 0.08 & 0\\
AGES~J114148+194538 & 11:41:48.0 & 0.7 & 19:45:38& 10 & 7811 \pm 6 & 138  \pm 12 & 201 \pm 18 & 6.0 \pm 0.7 & 0.73 \pm 0.09 & 0\\
AGES~J114204+194536 & 11:42:04.0 & 0.7 & 19:45:36& 10 & 11059 \pm 7 & 172 \pm 15 & 210 \pm 22 & 3.4 \pm 0.6 & 0.35 \pm 0.08 & 1\\
AGES~J114209+201924 & 11:42:09.3 & 0.7 & 20:19:24& 10 & 5764 \pm 5 & 319  \pm 9 & 337 \pm 14  & 5.9 \pm 0.9 & 1.25 \pm 0.18 & 0\\
AGES~J114211+195827 & 11:42:11.7 & 0.7 & 19:58:27& 11 & 7774 \pm 6 & 145  \pm 12 & 246 \pm 18 & 6.8 \pm 0.6 & 0.88 \pm 0.09 & 0\\
AGES~J114216+200302 & 11:42:16.0 & 1.5 & 20:03:02& 19 & 6087 \pm 5 & 112  \pm 9 & 173 \pm 14  & 6.7 \pm 0.6 & 0.74 \pm 0.08 & 0\\
AGES~J114224+200710 & 11:42:24.4 & 0.7 & 20:07:10& 10 & 5979 \pm 2 & 334  \pm 3 & 353 \pm 5   & 17.8 \pm 1.0 & 4.30 \pm 0.20 & 0\\
AGES~J114239+201150 & 11:42:39.6 & 0.8 & 20:11:50& 10 & 5993 \pm 8 & 169  \pm 16 & 263 \pm 23 & 4.6 \pm 0.6 & 0.56 \pm 0.08 & 0\\
AGES~J114243+200143 & 11:42:43.8 & 0.7 & 20:01:43& 11 & 6338 \pm 3 & 300  \pm 6 & 307 \pm 10  & 3.0 \pm 0.5 & 0.46 \pm 0.08 & 0\\
AGES~J114255+195734 & 11:42:55.9 & 0.7 & 19:57:34& 10 & 7264 \pm 7 & 186  \pm 13 & 292 \pm 20 & 7.5 \pm 0.7 & 1.17 \pm 0.11 & 0\\
AGES~J114308+194210 & 11:43:08.1 & 0.7 & 19:42:10& 10 & 12952 \pm 5 & 313 \pm 10 & 371 \pm 15 & 5.9 \pm 0.6 & 1.32 \pm 0.11 & 0\\
AGES~J114311+200032 & 11:43:11.8 & 0.7 & 20:00:32& 11 & 7085 \pm 7 & 65   \pm 14 & 112 \pm 21 & 4.2 \pm 0.6 & 0.25 \pm 0.06 & 2\\
AGES~J114312+193655 & 11:43:12.0 & 0.7 & 19:36:55& 10 & 6247 \pm 4 & 56   \pm 8 & 93 \pm 12   & 7.2 \pm 0.7 & 0.36 \pm 0.06 & 0\\
AGES~J114331+193747 & 11:43:31.6 & 0.7 & 19:37:47& 10 & 13131 \pm 4 & 274 \pm 9 & 341 \pm 13  & 8.2 \pm 0.6 & 1.97 \pm 0.13 & 0\\
AGES~J114347+195820 & 11:43:47.1 & 0.7 & 19:58:20& 10 & 6904 \pm 8 & 165  \pm 16 & 547 \pm 25 & 13.2 \pm 0.8 & 2.65 \pm 0.15 & 2\\
AGES~J114358+200430 & 11:43:58.5 & 0.7 & 20:04:30& 10 & 7372 \pm 2 & 258  \pm 4 & 273 \pm 6   & 12.3 \pm 0.9 & 2.53 \pm 0.16 & 0\\
AGES~J114402+202313 & 11:44:02.5 & 0.8 & 20:23:13& 11 & 11843 \pm 6 & 243 \pm 11 & 274 \pm 17 & 6.4 \pm 1.0 & 0.72 \pm 0.13 & 0\\
AGES~J114416+201309 & 11:44:16.4 & 0.8 & 20:13:09& 12 & 6370 \pm 6 & 302  \pm 12 & 325 \pm 18 & 3.3 \pm 0.6 & 0.55 \pm 0.10 & 0\\
AGES~J114435+192841 & 11:44:35.6 & 0.7 & 19:28:41& 10 & 2765 \pm 2 & 133  \pm 3 & 144 \pm 5   & 8.9 \pm 0.7 & 1.05 \pm 0.10 & 0\\
AGES~J114447+200737 & 11:44:47.7 & 0.7 & 20:07:37& 11 & 6583 \pm 3 & 392  \pm 6 & 408 \pm 9   & 5.1 \pm 0.6 & 0.67 \pm 0.08 & 0\\
AGES~J114451+192723 & 11:44:51.8 & 0.7 & 19:27:23& 10 & 5478 \pm 8 & 374  \pm 15 & 456 \pm 23 & 5.2 \pm 0.7 & 1.25 \pm 0.13 & 0\\
AGES~J114452+194635 & 11:44:52.9 & 0.7 & 19:46:35& 10 & 8225 \pm 8 & 261  \pm 16 & 425 \pm 25 & 8.6 \pm 0.8 & 1.46 \pm 0.14 & 0\\
AGES~J114503+195832 & 11:45:03.2 & 0.7 & 19:58:32& 10 & 5091 \pm 2 & 465  \pm 4 & 494 \pm 6   & 18.8 \pm 1.1 & 6.37 \pm 0.26 & 0\\
AGES~J114508+200837 & 11:45:08.5 & 0.7 & 20:08:37& 11 & 3724 \pm 2 & 74   \pm 5 & 95 \pm 7    & 9.6 \pm 0.8 & 0.62 \pm 0.08 & 0\\
AGES~J114543+200147 & 11:45:43.5 & 0.7 & 20:01:47& 10 & 5322 \pm 2 & 45   \pm 4 & 80 \pm 6    & 38.0 \pm 2.0 & 1.56 \pm 0.16 & 0\\
AGES~J114545+194210 & 11:45:45.9 & 0.8 & 19:42:10& 12 & 6182 \pm 7 & 154  \pm 14 & 183 \pm 21 & 3.1 \pm 0.6 & 0.33 \pm 0.08 & 1\\
AGES~J114612+202403 & 11:46:12.4 & 0.7 & 20:24:03& 11 & 7026 \pm 3 & 595  \pm 7 & 610 \pm 10  & 9.4 \pm 1.2 & 2.39 \pm 0.24 & 2\\
AGES~J114625+193024 & 11:46:25.5 & 0.7 & 19:30:24& 10 & 2189 \pm 3 & 103  \pm 5 & 121 \pm 8   & 6.0 \pm 0.6 & 0.54 \pm 0.07 & 0\\
AGES~J114638+194854 & 11:46:38.1 & 0.9 & 19:48:54& 14 & 5416 \pm 9 & 97   \pm 19 & 149 \pm 28 & 2.7 \pm 0.5 & 0.21 \pm 0.06 & 1\\
AGES~J114646+191906 & 11:46:46.8 & 0.7 & 19:19:06& 11 & 14644 \pm 9 & 173 \pm 18 & 277 \pm 28 & 7.2 \pm 1.0 & 0.96 \pm 0.14 & 0\\
AGES~J114724+192324 & 11:47:24.1 & 0.7 & 19:23:24& 10 & 5854 \pm 4 & 159  \pm 7 & 175 \pm 11  & 4.8 \pm 0.6 & 0.52 \pm 0.09 & 0\\
AGES~J114727+194906 & 11:47:27.4 & 0.8 & 19:49:06& 11 & 5869 \pm 7 & 60   \pm 15 & 99 \pm 22  & 3.6 \pm 0.6 & 0.19 \pm 0.06 & 1\\
AGES~J114730+195353 & 11:47:30.0 & 0.8 & 19:53:53& 14 & 5759 \pm 5 & 261  \pm 11 & 282 \pm 16 & 3.7 \pm 0.6 & 0.41 \pm 0.08 & 1\\
AGES~J114739+195621 & 11:47:39.7 & 0.7 & 19:56:21& 10 & 6173 \pm 2 & 386  \pm 4 & 396 \pm 6   & 7.1 \pm 0.7 & 1.29 \pm 0.12 & 0\\
AGES~J114739+200924 & 11:47:39.2 & 0.7 & 20:09:24& 10 & 5316 \pm 5 & 124  \pm 11 & 156 \pm 16 & 3.7 \pm 0.5 & 0.37 \pm 0.07 & 0\\
AGES~J114809+192109 & 11:48:09.1 & 2.4 & 19:21:09& 18 & 11252 \pm 5 & 171 \pm 9 & 179 \pm 14  & 3.3 \pm 0.8 & 0.34 \pm 0.10 & 1\\
AGES~J114823+194244 & 11:48:23.1 & 0.7 & 19:42:44& 10 & 5079 \pm 7 & 82   \pm 14 & 155 \pm 21 & 4.5 \pm 0.5 & 0.25 \pm 0.05 & 1\\
AGES~J114824+202331 & 11:48:24.3 & 0.9 & 20:23:31& 17 & 6750 \pm 8 & 80   \pm 15 & 143 \pm 23 & 9.8 \pm 1.4 & 0.64 \pm 0.14 & 0\\
AGES~J114829+194529 & 11:48:29.3 & 1.1 & 19:45:29& 11 & 5586 \pm 4 & 121  \pm 8 & 144 \pm 13  & 5.7 \pm 0.8 & 0.41 \pm 0.08 & 0\\
AGES~J114956+192840 & 11:49:56.7 & 0.8 & 19:28:40& 11 & 6139 \pm 4 & 67   \pm 7 & 82 \pm 11   & 3.9 \pm 0.5 & 0.23 \pm 0.05 & 1\\
AGES~J115030+200505 & 11:50:30.0 & 0.8 & 20:05:05& 11 & 6021 \pm 7 & 161  \pm 14 & 185 \pm 21 & 2.9 \pm 0.6 & 0.27 \pm 0.08 & 1\\
AGES~J115034+194301 & 11:50:34.4 & 0.7 & 19:43:01& 10 & 6633 \pm 7 & 88   \pm 15 & 162 \pm 22 & 5.2 \pm 0.7 & 0.38 \pm 0.07 & 0\\
AGES~J115052+193102 & 11:50:52.7 & 0.7 & 19:31:02& 11 & 11829 \pm 8 & 270 \pm 16 & 336 \pm 24 & 3.7 \pm 0.5 & 0.67 \pm 0.09 & 0\\
AGES~J115101+193728 & 11:51:01.0 & 0.7 & 19:37:28& 11 & 11626 \pm 4 & 247 \pm 8 & 268 \pm 12  & 4.3 \pm 0.5 & 0.53 \pm 0.08 & 0\\
AGES~J115101+202355 & 11:51:01.8 & 0.7 & 20:23:55& 10 & 6453 \pm 2 & 269  \pm 4 & 284 \pm 5   & 21.8 \pm 1.5 & 3.45 \pm 0.24 & 0\\
AGES~J115120+192931 & 11:51:20.4 & 0.8 & 19:29:31& 10 & 11971 \pm 4 & 201 \pm 9 & 226 \pm 13  & 4.2 \pm 0.5 & 0.42 \pm 0.07 & 0\\
AGES~J115147+192057 & 11:51:47.4 & 0.7 & 19:20:57& 11 & 6931 \pm 3 & 343  \pm 6 & 354 \pm 9   & 5.1 \pm 0.7 & 0.61 \pm 0.09 & 2\\
AGES~J115150+194720 & 11:51:50.7 & 1.0 & 19:47:20& 14 & 6361 \pm 4 & 102  \pm 9 & 119 \pm 13  & 3.2 \pm 0.5 & 0.23 \pm 0.06 & 1\\
AGES~J115156+193838 & 11:51:56.1 & 0.7 & 19:38:38& 10 & 6060 \pm 2 & 187  \pm 5 & 212 \pm 7   & 11.7 \pm 0.8 & 1.58 \pm 0.12 & 0\\
AGES~J115202+193423 & 11:52:02.3 & 0.7 & 19:34:23& 10 & 11989 \pm 5 & 163 \pm 10 & 186 \pm 15 & 3.3 \pm 0.5 & 0.33 \pm 0.07 & 0\\
AGES~J115223+201943 & 11:52:23.5 & 0.7 & 20:19:43& 12 & 6491 \pm 5 & 128  \pm 10 & 180 \pm 15 & 8.6 \pm 0.9 & 0.95 \pm 0.12 & 0\\
AGES~J115227+200821 & 11:52:27.9 & 0.7 & 20:08:21& 10 & 16085 \pm 7 & 140 \pm 14 & 167 \pm 20 & 3.2 \pm 0.6 & 0.35 \pm 0.08 & 1\\
AGES~J115250+201423 & 11:52:50.5 & 0.7 & 20:14:23& 10 & 14678 \pm 3 & 186 \pm 6 & 218 \pm 10  & 10.2 \pm 0.9 & 1.64 \pm 0.14 & 0\\
AGES~J115318+200622 & 11:53:18.0 & 0.7 & 20:06:22& 10 & 6215 \pm 3 & 112  \pm 5 & 163 \pm 8   & 19.7 \pm 1.1 & 2.01 \pm 0.14 & 0\\
AGES~J115349+194505 & 11:53:49.8 & 0.7 & 19:45:05& 10 & 12047 \pm 6 & 447 \pm 12 & 493 \pm 17 & 6.0 \pm 0.8 & 1.28 \pm 0.14 & 0\\
AGES~J115356+201043 & 11:53:56.2 & 0.8 & 20:10:43& 11 & 15210 \pm 4 & 228 \pm 8 & 236 \pm 11  & 3.7 \pm 0.7 & 0.49 \pm 0.11 & 1-2\\
AGES~J115358+195734 & 11:53:58.9 & 0.7 & 19:57:34& 11 & 15336 \pm 9 & 125 \pm 17 & 193 \pm 26 & 4.0 \pm 0.6 & 0.43 \pm 0.08 & 2\\
AGES~J115414+200138 & 11:54:14.1 & 0.7 & 20:01:38& 11 & 6208 \pm 3 & 225  \pm 7 & 249 \pm 10  & 10.0 \pm 1.0 & 1.50 \pm 0.16 & 0\\
\noalign{\smallskip}
\hline
\end{array}
\]
\end{table*}

\begin{table*}
\tiny
\caption{{HI parameters for the 22 sources observed with Lwide and confirmed.}
\label{lwidetab}} 
\[
\begin{array}{cccccccc}
\hline
\noalign{\smallskip}
HI-ID  &  V & W50  & W20  & F_{tot}   & rms\\
              & \rm km\,s^{-1}    & \rm  km\,s^{-1}  & \rm  km\,s^{-1}    & \rm Jy~km\,s^{-1}   & mJy   \\
\noalign{\smallskip}
\hline
\noalign{\smallskip}
AGES~J113508+201210 &16144   \pm  6  &  99 \pm 12  &  128 \pm 18  &  0.37 \pm 0.08 & 0.77\\
AGES~J113524+200009 & 9556   \pm  7  & 303 \pm 13  &  340 \pm 20  &  0.40 \pm 0.07 & 0.52\\
AGES~J113801+200106 &19086   \pm 11  & 243 \pm 22  &  372 \pm 33  &  0.89 \pm 0.12 & 0.80\\
AGES~J113835+195550 & 13198  \pm  5  & 286 \pm  9  &  294 \pm 14  &  0.52 \pm 0.12  & 0.80\\
AGES~J113904+194129 & 16917  \pm 10  &  54 \pm 19  &  113 \pm 29  &  0.22 \pm 0.07 & 0.70\\
AGES~J113926+194459 & 17186  \pm  7  & 167 \pm 15  &  203 \pm 22  &  0.45 \pm 0.10 & 0.80\\
AGES~J113942+200817 &16915   \pm  7  &  25 \pm 14  &   87 \pm 21  &  0.13 \pm 0.05 & 0.80\\
AGES~J114113+193240 & 10008  \pm  9  & 256 \pm 19  &  297 \pm 28  &  0.45 \pm 0.11 & 0.75\\
AGES~J114129+200550 & 14584  \pm  7  & 161 \pm 14  &  190 \pm 21  &  0.22 \pm 0.06 & 0.54\\
AGES~J114131+201912 &  3506  \pm  6  &  68 \pm 12  &   98 \pm 17  &  0.22 \pm 0.06  & 0.80\\
AGES~J114204+194536 & 11066  \pm 15  & 134 \pm 30  &  225 \pm 45  &  0.22 \pm 0.07 & 0.55\\
AGES~J114545+194210 &  6168  \pm 10  & 147 \pm 20  &  181 \pm 30  &  0.26 \pm 0.08 & 0.58\\
AGES~J114638+194854 &  5421  \pm 12  & 113 \pm 23  &  184 \pm 34  &  0.18 \pm 0.06 & 0.55\\
AGES~J114727+194906 &  5864  \pm  7  &  86 \pm 13  &  113 \pm 19  &  0.19 \pm 0.05 & 0.52\\
AGES~J114730+195353 &  5765  \pm 13  & 176 \pm 26  &  287 \pm 38  &  0.36 \pm 0.08 & 0.52\\
AGES~J114809+192109 & 11249  \pm  9  &  71 \pm 18  &  128 \pm 26  &  0.18 \pm 0.06 & 0.72\\
AGES~J114823+194244 &  5078  \pm  8  &  67 \pm 16  &  102 \pm 24  &  0.24 \pm 0.08 & 0.75\\
AGES~J114956+192840 &  6136  \pm  4  &  76 \pm  8  &   87 \pm 11  &  0.09 \pm 0.04 & 0.52\\
AGES~J115030+200505 &	6022 \pm  4  & 177 \pm  8  &  247 \pm 12  &  0.28 \pm 0.02 & 0.60\\
AGES~J115150+194720 &	6357 \pm  4  & 104 \pm  8  &  115 \pm 12  &  0.25 \pm 0.07 & 0.76\\
AGES~J115227+200821 &  16091 \pm  7  & 164 \pm 14  &  211 \pm 21  &  0.34 \pm 0.07 & 0.56\\
AGES~J115356+201043 & 15197  \pm  5  & 228 \pm 10  &  248 \pm 14  &  0.47 \pm 0.09 & 0.80\\
\noalign{\smallskip}
\hline
\end{array}
\]
\end{table*}

\begin{table*}
\tiny
\caption{{List of optical counterparts for the 100 HI sources in A1367}
\label{optcounttab}} 
\[
\begin{array}{ccccc}
\hline
\noalign{\smallskip}
HI-ID  &  Optical-ID & R.A.  & Dec.  & Optical~Velocity   \\
              &                  & \rm (J.2000) & \rm (J.2000)  & \rm km\,s^{-1}     \\
\noalign{\smallskip}
\hline
\noalign{\smallskip}
AGES~J113444+201217  &  MAPS-NGPO\_434\_0001768    &  11:34:46.01 & 20:12:17.2   &  9652^{*}   \\
AGES~J113508+201210  &  SDSSJ113508.79+201216.1  &  11:35:08.79 & 20:12:16.1   & --	 \\
AGES~J113523+201823  &  KUG1132+205              &  11:35:23.13 & 20:18:42.4   &  10451  \\
AGES~J113524+200009  &  MAPS-NGPO\_375\_1208433    &  11:35:27.77 & 20:00:2.7    & --	 \\
AGES~J113544+195120  &  SDSSJ113544.31+195114.0  &  11:35:44.31 &  19:51:14       & --	 \\
AGES~J113553+201308  &  KUG1133+204              &  11:35:54.44 & 20:13:19.9   &  1097^{*}   \\
AGES~J113614+195910  &  --                       &   --         & --	       & --	 \\
AGES~J113626+195102  &  -- 			    &   --      &    --        & --	 \\
AGES~J113641+200622  &  MAPS-NGPO\_434\_0002451    &  11:36:41.66 & 20:06:38.9   & --	 \\
AGES~J113653+200003  &  KUG1134+202A             &  11:36:54.24 & 19:59:49.9   &  6630   \\
AGES~J113657+195837  &  UGC06583                 &  11:36:54.40 & 19:58:15.0   &  6191   \\
AGES~J113718+192129  &  KUG1134+196              &  11:37:17.60 & 19:21:36.0   &  14370  \\
AGES~J113736+200934  &  CGCG097-033              &  11:37:36.00 & 20:09:49.0   &  7736   \\
AGES~J113743+195559  &  SDSSJ113744.57+195556.7  &  11:37:44.57 & 19:55:56.7   & --	 \\
		     &  SDSSJ113744.21+195616.2  &  11:37:44.21 & 19:56:16.2   & --	 \\
		     &  SDSSJ113743.51+195555.4  &  11:37:43.51 & 19:55:55.4   & --	 \\
AGES~J113801+200106  &  SDSS J113804.47+200021.8 &  11:38:04.47 & 20:00:21.8   & --	 \\
	             &  SDSS J113804.46+200043.4 &  11:38:04.46 & 20:00:43.4   & --	 \\
AGES~J113804+195146  &  2MASXJ11380386+1951420   &  11:38:03.90 & 19:51:41.0   &  6196   \\
AGES~J113804+201441  &  MAPS-NGPO\_434\_0002980    &  11:38:00.73 & 20:14:32.3   & --	 \\
AGES~J113805+193250  &  2MASXJ11380463+1933020   &  11:38:05.00 & 19:33:01.0   &  17891  \\
AGES~J113828+195823  &  LSBCD571-03              &  11:38:28.16 & 19:58:50.1   &  6981   \\
AGES~J113835+195550  &  2MASXJ11383847+1954564   &  11:38:38.49 & 19:54:56.7   & --	 \\
AGES~J113842+200527  &  SDSSJ113845.65+200511.4  &  11:38:45.65 & 20:05:11.4   & --	 \\
AGES~J113844+200821  &  KUG1136+204              &  11:38:45.26 & 20:08:24.1   &  3131^{*}   \\
AGES~J113850+193622  &  KUG1136+198              &  11:38:51.00 & 19:36:04.0   &  6787   \\
AGES~J113904+194129  &  SDSSJ113906.75+194135.1  &  11:39:06.75 & 19:41:35.1   & --	 \\
AGES~J113910+193558  &  FGC1287                  &  11:39:10.95 & 19:35:06.0   &  6825   \\
AGES~J113913+195915  &  MAPS-NGPO\_434\_0016203    &  11:39:11.00 & 19:59:00.9   & --	 \\
AGES~J113922+193232  &  CGCG097-041              &  11:39:24.50 & 19:32:04.0   &  6778   \\
AGES~J113926+194459  &  KUG1136+200              &  11:39:28.71 & 19:44:12.1   & --	 \\
AGES~J113930+191706  &  MAPS-NGPO\_434\_0045678    &  11:39:31.78 & 19:17:24.9   & --	 \\
AGES~J113936+202020  &  2MASXJ11393700+2020039   &  11:39:37.00 & 20:20:03.9   & --	 \\
AGES~J113939+193524  &  SDSSLSB                  &  11:39:40.17 & 19:35:20.0   & --	 \\
AGES~J113942+200817  &  SDSSJ113945.23+200806.9  &  11:39:45.23 & 20:08:07.0   & --	 \\
AGES~J113945+195341  &  KUG1137+201A             &  11:39:44.85 & 19:53:57.8   & --	 \\
AGES~J113947+195559  &  UGC06625                 &  11:39:47.60 & 19:55:60.0   &  10964  \\
AGES~J113956+195955  &  KUG1137+202B             &  11:39:57.20 & 20:00:13.0   &  10098  \\
AGES~J114016+194715  &  [IBG2003]J114017+194718  &  11:40:17.26 & 19:47:19.1   & --	 \\
AGES~J114027+192429  &  KUG1137+196              &  11:40:26.15 & 19:24:51.9   &  3421   \\
AGES~J114039+195455  &   [IBC2002]J114038+195437 &  11:40:38.99 & 19:54:38.5   &  7784   \\
AGES~J114113+193240  &  KUG1138+198              &  11:41:13.43 & 19:32:21.7   &  10011  \\
AGES~J114118+200829  &  KUG 1138+204             &  11:41:17.45 & 20:08:33.4   &  14500  \\
AGES~J114129+200550  &  SDSSJ114130.18+200629.8  &  11:41:30.18 & 20:06:29.8   & --	 \\
                     &  SDSSJ114129.01+200550.4  &  11:41:29.01 & 20:05:50.4   & --	 \\
                     &  SDSSJ114129.74+200506.0  &  11:41:29.74 & 20:05:06.0   & --	 \\
AGES~J114131+201912  &  SDSSJ114134.75+201917.9  &  11:41:34.75 & 20:19:17.9   & --	 \\
                     &  SDSSJ114128.58+201803.8  &  11:41:28.58 & 20:18:03.8   & --	 \\
AGES~J114143+193536  &  MAPS-NGPO\_434\_0037838    &  11:41:43.22 & 19:35:26.2   & --	 \\
AGES~J114148+194538  &  [IBC2002]J114149+194605  &  11:41:49.83 & 19:46:04.4   &  7789   \\
AGES~J114204+194536  &  SDSSJ114205.74+194611.7  &  11:42:05.74 & 19:46:11.7   & --	 \\
AGES~J114209+201924  &  NGC3821                  &  11:42:09.09 & 20:18:56.6   &  5788   \\
AGES~J114211+195827  &  KUG1139+202  	         &  11:42:14.77 & 19:58:35.2   &  7815   \\
AGES~J114216+200302  &  KUG1139+203  	         &  11:42:15.66 & 20:02:55.5   &  6102   \\
AGES~J114224+200710  &  CGCG097-068  	         &  11:42:24.48 & 20:07:09.9   &  5974   \\
AGES~J114239+201150  &  --                       &   --         & --	       & --	 \\
AGES~J114243+200143  &  CGCG097-072  	         &  11:42:45.21 & 20:01:56.6   &  6332   \\
AGES~J114255+195734  &  CGCG097-073  	         &  11:42:56.46 & 19:57:58.4   &  7275   \\
AGES~J114308+194210  &  KUG1140+199  	         &  11:43:07.96 & 19:41:57.4   &  12988  \\
AGES~J114311+200032  &  CGCG097-079  	         &  11:43:13.32 & 20:00:17.5   &  7000   \\
AGES~J114312+193655  &  [IBC2002]J114313+193645  &  11:43:13.02 & 19:36:46.6   &  6121   \\
AGES~J114331+193747  &  CGCG097-083              &  11:43:31.22 & 19:37:40.6   &  13126  \\
AGES~J114347+195820  &  UGC06697                 &  11:43:49.11 & 19:58:06.2   &  6725   \\
AGES~J114358+200430  &  NGC3840                  &  11:43:59.01 & 20:04:37.3   &  7368   \\
AGES~J114402+202313  &  2MASXJ11435706+2022383   &  11:44:01.92 & 20:22:51.6   &  11962  \\
AGES~J114416+201309  &  CGCG097-102              &  11:44:17.22 & 20:13:23.9   &  6368   \\
AGES~J114435+192841  &  MAPS-NGPO\_434\_0047286    &  11:44:36.39 & 19:28:40.1   & --	 \\
AGES~J114447+200737  &  UGC06719                 &  11:44:47.12 & 20:07:30.2   &  6571   \\
AGES~J114451+192723  &  NGC3859                  &  11:44:52.24 & 19:27:15.2   &  5468   \\
AGES~J114452+194635  &  CGCG097-125              &  11:44:54.85 & 19:46:34.9   &  8271   \\
AGES~J114503+195832  &  NGC3861                  &  11:45:03.87 & 19:58:25.1   &  5085   \\
AGES~J114508+200837  &  [IBG2003] J114506+200849 &  11:45:06.40 & 20:08:49.2   &  3822   \\
AGES~J114543+200147  &  CGCG097-138              &  11:45:44.67 & 20:01:51.6   &  5313   \\
AGES~J114545+194210  &  [IBG2003]J114545+194130  &  11:45:45.27 & 19:41:30.7   &  6123   \\
AGES~J114612+202403  &  NGC3884                  &  11:46:12.19 & 20:23:29.9   &  6946   \\
AGES~J114625+193024  &  SDSSJ114626.45+193011.9  &  11:46:26.45 & 19:30:11.9   & --	 \\
AGES~J114638+194854  &  SDSSJ114636.47+194834.1  &  11:46:36.47 & 19:48:34.1   & --	 \\
                     &  SDSSJ114640.32+194820.8  &  11:46:40.32 & 19:48:20.8   & --	 \\
AGES~J114646+191906  &  2MASXJ11464700+1919462   &  11:46:47.03 & 19:19:46.2   & --	 \\
AGES~J114724+192324  &  MAPS-NGPO\_434\_0048296    &  11:47:25.08 & 19:23:31.4   & --	 \\
AGES~J114727+194906  &  SDSSJ114730.39+194859.0  &  11:47:30.39 & 19:48:59.0   & --	 \\
AGES~J114730+195353  &  2MASXJ11473099+1952201   &  11:47:30.99 & 19:52:20.8   &  5778   \\
AGES~J114739+195621  &  CGCG097-152              &  11:47:39.35 & 19:56:22.0   &  6166   \\
AGES~J114739+200924  &  SDSSJ114737.57+200900.6  &  11:47:37.57 & 20:09:00.6   & --	 \\
AGES~J114809+192109  &  --                       &   --         & --	       & --	 \\
AGES~J114823+194244  &  SDSSJ114825.21+194217.0  &  11:48:25.21 & 19:42:17.0   & --	 \\
AGES~J114824+202331  &  LSBCD571-02              &  11:48:26.40 & 20:23:19.0   &  6796   \\
AGES~J114829+194529  &  MAPS-NGPO\_434\_0031444    &  11:48:28.43 & 19:45:36.0   & --	 \\
AGES~J114956+192840  &  SDSSJ114956.75+192840.4  &  11:49:56.76 & 19:28:40.4   & --	 \\
                     &  SDSSJ114956.93+192830.7  &  11:49:56.93 & 19:28:30.7   & --	 \\
AGES~J115030+200505  &  MAPS-NGPO\_434\_0021239    &  11:50:29.99 & 20:05:08.2   & --	 \\
AGES~J115034+194301  &  SDSSJ115033.61+194250.1  &  11:50:33.61 & 19:42:50.1   & --	 \\
AGES~J115052+193102  &  KUG1148+197A             &  11:50:52.48 & 19:31:10.8   & --	 \\
AGES~J115101+193728  &  2MASXJ11510128+1937311   &  11:51:01.29 & 19:37:31.3   & --	 \\
AGES~J115101+202355  &  UGC06821                 &  11:51:01.12 & 20:23:57.3   &  6438   \\
AGES~J115120+192931  &  KUG1148+197              &  11:51:21.28 & 19:29:33.9   & --	 \\
AGES~J115147+192057  &  KUG1149+196              &  11:51:48.37 & 19:21:29.3   &  6932   \\
\noalign{\smallskip}
\hline
\end{array}
\]
*: new optical redshift obtained in Winter 2007 using the Loiano 1.5 meter telescope.
\end{table*}

\setcounter{table}{2}
\begin{table*}
\tiny
\caption{{Continue.}} 
\[
\begin{array}{ccccc}
\hline
\noalign{\smallskip}
HI-ID  &  Optical-ID & R.A.  & Dec.  & Optical~Velocity   \\
              &                  & \rm (J.2000) & \rm (J.2000)  & \rm km\,s^{-1}     \\
\noalign{\smallskip}
\hline
\noalign{\smallskip}
AGES~J115150+194720  &  SDSSJ115150.33+194702.6  &  11:51:50.33 & 19:47:02.7   & --	 \\
                     &  SDSSJ115150.87+194613.2  &  11:51:50.87 & 19:46:13.2   & --	 \\
AGES~J115156+193838  &  KUG1149+199              &  11:51:55.65 & 19:38:45.4   &  6134^{*}   \\
AGES~J115202+193423  &  SDSSJ115203.86+193425.5  &  11:52:03.86 & 19:34:25.5   & --	 \\
AGES~J115223+201943  &  KUG1149+206              &  11:52:24.44 & 20:19:28.8   &  6491   \\
AGES~J115227+200821  &  2MASXJ11522887+2007597   &  11:52:28.85 & 20:07:59.5   & --	 \\
AGES~J115250+201423  &  KUG1150+205              &  11:52:51.04 & 20:14:21.9   &  14679  \\
AGES~J115318+200622  &  KUG1150+203              &  11:53:18.28 & 20:06:26.6   &  6061^{*}   \\
AGES~J115349+194505  &  CGCG097-176              &  11:53:50.34 & 19:45:01.5   &  12005  \\
AGES~J115356+201043  &  MAPS-NGPO\_376\_4233155    &  11:53:55.94 & 20:10:55.6   & --	 \\
AGES~J115358+195734  &  KUG1151+202              &  11:53:59.89 & 19:57:27.1   &  15489  \\
AGES~J115414+200138  &  KUG1151+203              &  11:54:13.94 & 20:01:39.0   &  6187   \\
\noalign{\smallskip}
\hline
\end{array}
\]
*: new optical redshift obtained in Winter 2007 using the Loiano 1.5 meter telescope.
\end{table*}

\begin{table*}
\tiny
\caption{{SDSS-Blue galaxies undetected in HI.}
\label{nodettab}} 
\[
\begin{array}{cccccccccccc}
\hline
\noalign{\smallskip}
 R.A.               & Dec.       & Name. & u & g & r & i & z & g-i  & V & R_{iso}(r) & R_{iso}(H\alpha) \\
\rm (J.2000)  & (J.2000) &             & mag& mag & mag & mag & mag & mag &  \rm km\,s^{-1} & arcsec  & arcsec   \\
\noalign{\smallskip}
\hline
\noalign{\smallskip}
11:42:18.12 & +29:50:15.9& KUG 1139+201 		&  17.15 & 15.88 & 15.40 & 15.12 &14.98 & 0.76 & 6476  & 11.6 & 3.0 \\
11:43:47.76 & +20:21:48.0& KUG 1141+206A		&  17.59 & 16.12 & 15.66 & 15.43 &15.25 & 0.71 & 6722  & 15.1 & 6.8 \\
11:43:48.90 & +20:14:54.0& KUG 1141+205 		&  17.33 & 16.35 & 15.90 & 15.78 &15.60 & 0.57 & 6103  & 8.0   & 4.8 \\
11:43:49.88 & +19:58:34.8& CGCG 97-087N 		&  18.04 & 16.78 & 16.49 & 16.18 &16.01 & 0.60 & 7542  & 19.0 & 6.0 \\
11:43:58.23 & +20:11:07.9& CGCG 97-092  		&  16.46 & 15.39 & 14.95 & 14.70 &14.52 & 0.70 & 6487  & -- & --\\
11:44:01.94 & +19:47:03.9& CGCG 97-093  		&  16.37 & 15.47 & 15.18 & 15.05 &14.98 & 0.42 & 4909  & 24.0 & 9.1 \\
11:44:13.90 & +19:20:11.3& MAPS-NGP O\_434\_0047177	&  17.48 & 16.96 & 16.79 & 16.73 &16.66 & 0.23 & 5852 & 10.3   & 8.6 \\
11:44:54.56 & +20:01:01.4& KUG 1142+202A		&  17.85 & 16.65 & 16.20 & 15.93 &15.73 & 0.73 & 7646  & 10.8 & 3.2 \\
\noalign{\smallskip}
\hline
\end{array}
\]
\end{table*}

\end{document}